\documentclass[12pt,a4paper,twoside,fleqn]{article}
\usepackage{graphicx}
\usepackage{cite}

\def\be{\begin{equation}}
\def\ee{\end{equation}}
\def\bea{\begin{eqnarray}}
\def\eea{\end{eqnarray}}

\def\bbuildrel#1_#2^#3{\mathrel{\mathop{\kern 0pt#1}\limits_{#2}^{#3}}}
\def\slash#1{\setbox0=\hbox{$#1$}#1\hskip-\wd0\dimen0=5pt\advance
       \dimen0 by-\ht0\advance\dimen0 by\dp0\lower0.5\dimen0\hbox
         to\wd0{\hss\sl/\/\hss}}

\newcommand{\scs}{\scriptscriptstyle}
\newcommand{\f}{\frac}
\newcommand{\al}{\alpha_s}

\newcommand{\me}[1]{\langle#1\rangle}

\def\theequation{\thesection.\arabic{equation}}
\newcommand{\newsection}[1]{\section{#1}\setcounter{equation}{0}}
\newcommand{\newappendix}[1]{\section*{#1}\setcounter{equation}{0}}

\begin{document}
\begin{titlepage}

\begin{flushright}
  {\bf CERN-TH/2001-089\\
       IFT-11/2001\\
       hep-ph/0104034\\  
Nucl. Phys. B611 (2001) 338$^\star$} \\[1cm]
\end{flushright}

\begin{center}

\setlength {\baselineskip}{0.3in} 
{\bf\Large Quark mass effects in $\bar{B} \to X_s \gamma$}\\[2cm]

\setlength {\baselineskip}{0.2in}
{\large  Paolo Gambino$^{^{1}}$ and  Miko{\l}aj Misiak$^{^{1,\,2}}$}\\[5mm]

$^{^{1}}${\it Theory Division, CERN, CH-1211 Geneva 23, Switzerland}\\[3mm]

$^{^{2}}${\it Institute of Theoretical Physics, Warsaw University,\\
                 Ho\.za 69, PL-00-681 Warsaw, Poland}\\[2cm] 

{\bf Abstract}\\
\end{center} 
\setlength{\baselineskip}{0.2in} 

The charm-loop contribution to 
$\bar{B} \to X_s \gamma$ 
is found to be numerically dominant and very stable under logarithmic
QCD corrections. The strong enhancement of the branching ratio by QCD
logarithms is mainly due to the $b$-quark mass evolution in the
top-quark sector. These observations allow us to achieve better
control over residual scale-dependence at the next-to-leading order.
Furthermore, we observe that the sensitivity of the matrix element
$\me{X_s \gamma | (\bar{s}c)_{V-A}(\bar{c}b)_{V-A} | b}$ to $m_c/m_b$ 
is the source of a sizeable uncertainty that has not been properly taken
into account in previous analyses. Replacing 
$m_c^{\rm pole}/m_b^{\rm pole}$ 
in this matrix element by the more appropriate
$m_c^{\overline{\rm MS}}(\mu)/m_b^{\rm pole}$ with $\mu \in [m_c,m_b]$
causes an 11\% 
enhancement of the SM prediction for BR$[\bar{B} \to X_s \gamma]$. 
For $E_{\gamma} > 1.6$~GeV in the $\bar{B}$-meson rest frame,
we find
BR$[\bar{B} \to X_s \gamma]_{E_{\gamma} > 1.6~{\rm GeV}} = (3.60 \pm
0.30)\times 10^{-4}$. The difference between our result and the
current experimental world average is consistent with zero at the
level of $1\sigma$. We also discuss the implementation of new
physics effects in our calculation. The Two--Higgs--Doublet--Model~II
with a charged Higgs boson lighter than 350~GeV is found to be
strongly disfavoured.

\vspace{1cm}

\setlength {\baselineskip}{0.2in}
\noindent \underline{\hspace{2in}}\\ 
$^\star$ {\footnotesize Comparison with experiment is updated in the
  present hep-ph version.  In particular, the published results of
  CLEO \cite{CLEO} are taken into account.}

\end{titlepage} 

\setlength{\baselineskip}{0.23in}

\newsection{Introduction}
\label{sec:intro}

Strong constraints on new physics from $\bar{B} \to X_s \gamma$
\cite{CDGG98,DGG00,MPR98} crucially depend on theoretical
uncertainties in the Standard Model prediction for this decay.\footnote{
  $\bar{B}$ denotes either $\bar{B}^0_d$ or $B^-$, while $X_s$ stands for
  $S=-1$ hadronic states containing no charmed particles.}
 This becomes more and more transparent with the progress in 
  experimental accuracy. The current experimental results are
\bea
{\rm BR}[ \bar{B} \to X_s \gamma] &=& 
\left[ 3.21 \pm 0.43_{\rm stat} \pm 0.27_{\rm sys} 
~\left({}^{+0.18}_{-0.10}\right)_{\rm theory} \right] \times 10^{-4} 
\hspace{17mm} \mbox{(CLEO \cite{CLEO}),} \nonumber\\
{\rm BR}[ \bar{B} \to X_s \gamma] &=& 
\left[ 3.36 \pm 0.53_{\rm stat} \pm 0.42_{\rm sys} 
\pm 0.54_{\rm theory} \right] \times 10^{-4} 
\hspace{2cm} \mbox{(BELLE \cite{BELLE}),} \nonumber\\
{\rm BR}[ b \to s \gamma] &=& 
(3.11 \pm 0.80_{\rm stat} \pm 0.72_{\rm sys} ) \times 10^{-4} 
\hspace{4cm} \mbox{(ALEPH \cite{ALEPH}).} \nonumber
\eea
The weighted average for BR$_\gamma\equiv$BR$[ \bar{B} \to X_s \gamma]$
is therefore\footnote{
  Statistical errors in the ALEPH measurement of $b \to s \gamma$ are
  much larger than expected differences among weak radiative branching
  ratios of the included $b$-hadrons.}
\be \label{main.exp}
{\rm BR}_\gamma^{\rm exp} = (3.23 \pm 0.42 ) \times 10^{-4}
\ee
with an error of around 13\%.  Improved results from CLEO and BELLE,
as well as new results from BABAR, are expected soon. Thus, all efforts
should be made, on the theoretical side, to reduce the uncertainty 
significantly below 10\%.

There are three sources of uncertainties in the theoretical
prediction: parametric, non-perturbative and perturbative. The most
important parametric ones in our analysis are due to $\al(M_Z)$ and
the HQET parameter $\lambda_1$. The latter parameter occurs in the
calculation of the phase-space factor of the semileptonic branching
ratio that is conventionally used for normalization.

All the non-perturbative effects, except for the calculated
$\Lambda^2/m_{c,b}^2$ corrections \cite{1.ov.mc,1.ov.mb}, sum up to the
non-perturbative uncertainty. No satisfactory quantitative estimate of
those effects is available, but they are believed to be well below
10\% when the photon energy cut--off is above 1~GeV \cite{KLP95} and
below 1.9~GeV \cite{KN99} in the ${\bar B}$-meson rest frame. The
cut--offs imposed most recently by CLEO and BELLE are 2.0 and 2.1 GeV,
respectively, in the $\Upsilon(4S)$ frame.

As far as the perturbative uncertainties are concerned, they were
dramatically reduced 4 years ago, after the completion of NLO QCD
calculations \cite{GHW96,CMM97,AG95,P96,MM95,match2}. A further improvement
came from electroweak corrections \cite{KN99,CM98,BM00,GH00}. The
unknown NNLO QCD corrections were estimated at the level of $\pm 7\%$
in ref.~\cite{KN99}, where it was pointed out that certain
scale-dependence cancellations at NLO seem accidental.

However, one source of  perturbative uncertainty was  not 
properly taken into account in the previous analyses \cite{KN99,
CMM97, GH00, BKP97}. It is related to the question of the  definitions of
$m_c$ and $m_b$ that should be used in the matrix element
$\me{P_2} \equiv \me{X_s \gamma | (\bar{s}c)_{V-A}(\bar{c}b)_{V-A} | b}$.
This matrix element is non-vanishing at two loops only, so the
renormalization scheme for $m_c$ and $m_b$ is an NNLO issue (unless
$\ln\f{m_b}{m_c}$ is treated as a large logarithm).  However, this
problem is numerically very important because of the sensitivity of
$\me{P_2}$ to $m_c/m_b$.  Changing $m_c/m_b$ in $\me{P_2}$ from ~$0.29
\pm 0.02$~ to ~$0.22 \pm 0.04$,~ i.e. from
$m_c^{\rm pole}/m_b^{\rm pole}$ to 
$m_c^{\overline{\rm MS}}(\mu)/m_b^{\rm pole}$ (with $\mu \in [m_c,m_b]$)
causes an increase of BR$_\gamma$ by around 11\%.  What matters for
$\bar{B} \to X_s \gamma$ is mainly the real part of $\me{P_2}$, where
the charm quarks are usually off-shell, with a momentum scale set by
$m_b^{\rm pole}$ (or some sizeable fraction of it). Therefore, as we
shall argue in appendix D, the choice of
$m_c^{\overline{\rm MS}}(\mu)/m_b^{\rm pole}$ with $\mu \in [m_c,m_b]$
seems more reasonable than $m_c^{\rm pole}/m_b^{\rm pole}$.

Once $m_c^{\overline{\rm MS}}(\mu)/m_b^{\rm pole}$ with $\mu \in
[m_c,m_b]$ is used in $\me{P_2}$, the uncertainty in BR$_\gamma$
significantly increases.  This is due in part to a strong
scale-dependence of $m_c(\mu)$. Moreover, in all the previous
analyses, the $m_c$-dependence of $\me{P_2}$ cancelled partially against
that  of the semileptonic decay rate.  Once the different
nature of the charm mass in the two cases is appreciated, the
cancellation no longer  takes place.

In the present paper, we perform a reanalysis of $\bar{B} \to X_s
\gamma$, taking the above problems with $m_c/m_b$ into account. At the
same time, we make several improvements in the calculation, which
allows us to maintain the theoretical uncertainty at the level of
around $\sim\!10\%$. In particular, a careful calculation of the
semileptonic phase--space factor is performed, after expressing it in
terms of an observable for which the NNLO expressions are known.
Moreover, good control over the behaviour of QCD perturbation series
in $\bar{B} \to X_s \gamma$ is achieved by splitting the charm- and
top-quark-loop contributions to the decay amplitude.  The overall
factor of $m_b$ is frozen at the electroweak scale in the top
contribution to the effective vertex
~$m_b (\bar{s}_L \sigma^{\mu \nu} b_R) F_{\mu \nu}$.~
All the remaining factors of $m_b$ are expressed in terms of the
bottom ``1S mass''. As argued in ref.~\cite{HLM99}, expressing the
kinematical factors of $m_b$ in inclusive $B$-meson decay rates in
terms of the 1S mass improves the behaviour of QCD perturbation series
with respect to what would be obtained using $m_b^{\overline{\rm
MS}}(m_b)$ or $m_b^{\rm pole}$.  When such an approach is used, no
sizeable accidental cancellations of scale-dependence in the NLO
expressions for BR$_\gamma$ are observed any more.

Splitting the charm and top contributions to the amplitude allows us
to better understand the origin of the well-known factor of
$\;\sim\!\! 3$ enhancement of BR$_\gamma$ by QCD logarithms. When the
splitting is performed at LO, the charm contribution is found to be
extremely stable under QCD renormalization group evolution. The
logarithmic enhancement of the branching ratio appears to be almost
entirely due to the top-quark sector. It can be attributed to the
large anomalous dimension of the $b$-quark mass.

Our paper is organized as follows. In section \ref{sec:leading},
splitting of the charm and top contributions is performed at the
leading-logarithmic level, and $m_b(\mu)$ in the top sector is shown
to be the main source of large QCD logarithms. In section
\ref{sec:NLO}, the NLO formulae for BR$_\gamma$ are written in such a
way that the NNLO uncertainties can be conveniently controlled.
Section \ref{sec:num} is devoted to the numerical analysis, our final
result being given in eq.~(\ref{main.num}). In section
\ref{sec:newphys}, constraints on new physics are briefly
discussed. Section \ref{sec:concl} contains our conclusions.

Details on specific points of our calculation as well as the formulae
that are necessary to make our analysis self-contained are collected
in the appendices. In appendix A, our input parameters are
collected. Appendix B is devoted to a determination of the mass ratio
$m_b^{\overline{MS}}(m_t)/m_b^{1S}$. In appendix C, we calculate the
semileptonic phase-space factor. Appendix D contains the determination
of the ratio $m_c/m_b$ that should be used in $\me{P_2}$.  We also
give there the analytical dependence of this matrix element on
$m_c/m_b$~ \cite{GHW96}. In appendix E, the relevant bremsstrahlung
formulae are summarized.

\newsection{Leading-order considerations}
\label{sec:leading}

The resummation of large QCD logarithms in $B$ decays usually begins
by  decoupling  the heavy electroweak bosons and the top quark.  In
the resulting effective theory, flavour-changing interactions are
present only in operators $P_i$ of dimension$\,>\!4\;$. Their Wilson
coefficients $C_i(\mu)$ evolve according to the Renormalization
Group Equations (RGEs) from the matching scale $\mu_0$ down to the
scale $\mu_b$ where the matrix elements of $P_i$ are evaluated.

\begin{figure}[htb]
\vspace{5mm}
\hspace*{11mm}  $\gamma$ \hspace{36.5mm} $\gamma$ 
\hspace{41.5mm} $\gamma$ \hspace{36.5mm} $\gamma$ \\[7mm] 
\hspace*{0mm} $u,c,t$ \hspace{9mm}     $u,c,t$ \hspace{12mm} 
            $W^{\pm}$ \hspace{8mm}   $W^{\pm}$ \hspace{15mm}
              $u,c,t$ \hspace{9mm}     $u,c,t$ \hspace{14mm} 
          $G^{\pm}$ \hspace{8mm} $G^{\pm}$ \\[-18mm] 
\includegraphics[width=75mm,angle=0]{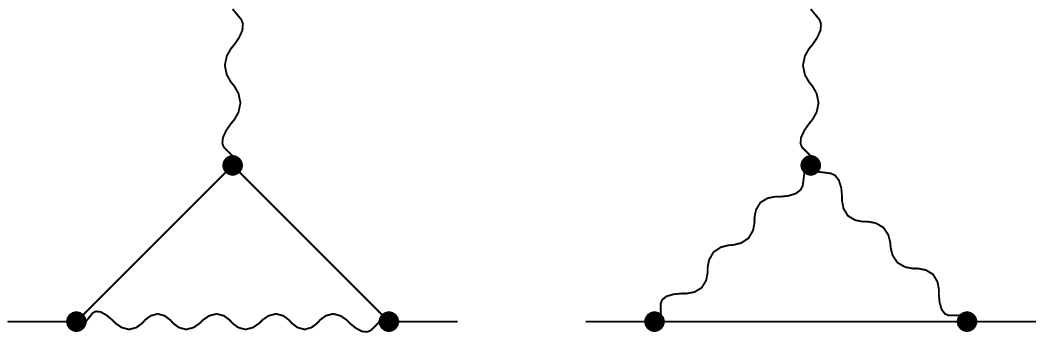}
\hspace{1cm}
\includegraphics[width=75mm,angle=0]{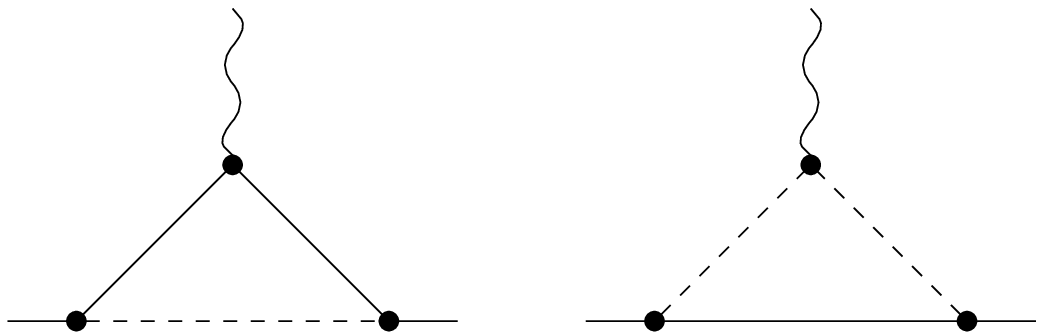}\\[-2mm]
$b$ \hspace{1cm}  $W^{\pm}$  \hspace{7mm} $s$ \hspace{6mm} 
$b$ \hspace{7mm}   $u,c,t$   \hspace{7mm} $s$ \hspace{11mm} 
$b$ \hspace{1cm} $G^{\pm}$ \hspace{8mm} $s$ \hspace{6.5mm} 
$b$ \hspace{7mm}   $u,c,t$   \hspace{7mm} $s$ \\[-1cm]
\begin{center}
\caption{\sf One-loop 1PI diagrams for $b \to s \gamma$ in the SM.
         There is no $W^{\pm}G^{\mp}\gamma$ coupling in the background-field gauge.}
\label{fig:one-loop}
\end{center}
\vspace{-1cm}
\end{figure}

In the case of $\bar{B} \to X_s \gamma$, the relevant operators $P_i$
read\footnote{
The CKM-suppressed $(\bar{s}u)(\bar{u}b)$ analogues of $P_1$ and $P_2$
are present in the effective theory, too. Their effects are included in
eq.~(\ref{c7eff0}) and in our NLO analysis in the following sections.}
\bea 
P_1  &=& (\bar{s}_L \gamma_{\mu} T^a c_L)        (\bar{c}_L \gamma^{\mu} T^a b_L),\nonumber\\
P_2  &=& (\bar{s}_L \gamma_{\mu}     c_L)        (\bar{c}_L \gamma^{\mu}     b_L),\nonumber\\
P_3  &=& (\bar{s}_L \gamma_{\mu}     b_L) \sum_q (\bar{q}   \gamma^{\mu}     q),  \nonumber\\
P_4  &=& (\bar{s}_L \gamma_{\mu} T^a b_L) \sum_q (\bar{q}   \gamma^{\mu} T^a q),  \nonumber\\
P_5  &=& (\bar{s}_L \gamma_{\mu}
                    \gamma_{\nu}
                    \gamma_{\rho}    b_L) \sum_q (\bar{q}   \gamma^{\mu} 
                                                            \gamma^{\nu}
                                                            \gamma^{\rho}    q),  \nonumber\\
P_6  &=& (\bar{s}_L \gamma_{\mu}
                    \gamma_{\nu}
                    \gamma_{\rho}T^a b_L) \sum_q (\bar{q}   \gamma^{\mu} 
                                                            \gamma^{\nu}
                                                            \gamma^{\rho}T^a q),  \nonumber\\
P_7  &=&  \f{e}{16 \pi^2} m_b(\mu) (\bar{s}_L \sigma^{\mu \nu}     b_R) F_{\mu \nu},
\nonumber\\[1mm]
P_8  &=&  \f{g}{16 \pi^2} m_b(\mu) (\bar{s}_L \sigma^{\mu \nu} T^a b_R) G_{\mu \nu}^a.
\label{ops}
\eea
In the leading logarithmic approximation, the $b \to s \gamma$
amplitude is proportional to the (effective) Wilson coefficient of the
operator $P_7$. The well-known \cite{BMMP94} expression for this
coefficient reads
\be      \label{c7eff0}
C^{(0)\rm eff}_7(\mu_b) = \eta^{\f{16}{23}} C^{(0)}_7(\mu_0) +
\f{8}{3} \left( \eta^{\f{14}{23}} - \eta^{\f{16}{23}}
\right) C^{(0)}_8(\mu_0)  + \sum_{i=1}^8 h_i \eta^{a_i}, 
\ee
where $\eta = \al(\mu_0)/\al(\mu_b)$ and
\be
h_i = \left( \begin{array}{cccccccc}
\f{626126}{272277} & -\f{56281}{51730} & -\f{3}{7} & -\f{1}{14} &
-0.6494 & -0.0380 & -0.0185 & -0.0057 \end{array} \right).
\ee
The powers $a_i$ are given in table~\ref{tab:mag}, in section
\ref{sec:NLO}. The coefficients $C^{(0)}_7(\mu_0)$ and
$C^{(0)}_8(\mu_0)$ are found from the one-loop electroweak diagrams
presented in fig.~\ref{fig:one-loop}.  It is sufficient to calculate the
1PI diagrams only.

Contributions from different internal quark flavours in those diagrams
can be separately matched onto gauge-invariant operators, even when
the calculation is performed off-shell.\footnote{
In an off-shell calculation, use of the background-field gauge 
is necessary to ensure the absence of gauge-non-invariant operators.} 
For the operator $P_7$ and its gluonic analogue $P_8$, each quark
flavour yields a UV-finite contribution that depends neither on the
renormalization scheme nor on the gauge-fixing parameter. 

When such a separation of flavours is made, and the CKM-suppressed
$u$-quark contribution is neglected, eq.~(\ref{c7eff0}) can be written
as
\be
C^{(0)\rm eff}_7(\mu_b) = X_c + X_t,
\ee
where the charm-quark contribution is given by
\be      \label{Xc}
X_c \;=\; -\f{23}{36} \eta^{\f{16}{23}} 
-\f{8}{9} \left( \eta^{\f{14}{23}} - \eta^{\f{16}{23}}\right) 
+ \sum_{i=1}^8 h_i \eta^{a_i},
\ee
and the top-quark one reads
\be      \label{Xt}
X_t \;=\; -\f{1}{2} A_0^t\left(\f{m_t^2}{M_W^2}\right) \eta^{\f{16}{23}} 
-\f{4}{3} F_0^t \left(\f{m_t^2}{M_W^2}\right) 
\left(\eta^{\f{14}{23}} - \eta^{\f{16}{23}}\right),
\ee
where
\be \label{A0.andF0}
\begin{array}{rcl}
A^t_0(x) &=& \f{-3x^3 + 2x^2}{2(x-1)^4} \ln x
+\f{-22x^3+153x^2-159x+46}{36(x-1)^3}, \\[3mm]
F^t_0(x) &=& \f{3x^2}{2(x-1)^4} \ln x
+\f{-5x^3+9x^2-30x+8}{12(x-1)^3}.\\[-3mm] \nonumber
\end{array}
\ee
The first two terms in $X_c$ (\ref{Xc}) are obtained from
eq.~(\ref{c7eff0}) by the following replacements: \linebreak \newpage \noindent
$C^{(0)}_7(\mu_0) \to -\f{23}{36}$ and $C^{(0)}_8(\mu_0) \to -\f{1}{3}$, which is
equivalent to including only charm contributions to the matching
conditions for the corresponding operators.  Analogously, only top
loops contribute to $X_t$. The last term in eq.~(\ref{c7eff0})
now appears in $X_c$, because it is entirely due to effects of charm
loops in the RGE evolution.  The splitting of charm and top is
performed at the level of SM Feynman diagrams, and the effective
theory is nothing but a technical tool for resumming large QCD
logarithms in gluonic corrections to those diagrams.
\begin{figure}[h]
\hspace*{13cm} $\eta$ \vspace{-7mm}
\begin{center}
\includegraphics[width=9cm,angle=0]{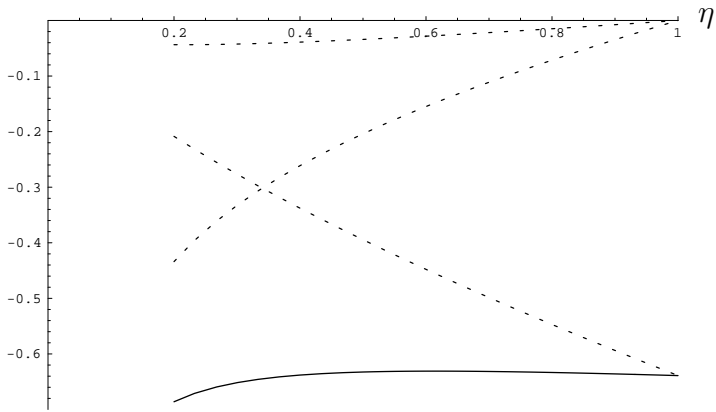}
\label{fig:eta-dep}
\end{center}
\caption{\sf $X_c$ as a function of $\eta$ (solid line), and its three components
          in eq.~(\ref{Xc}) (dashed lines).}
\end{figure}

$X_c$ is a function of $\eta$ that varies very slowly in the physically
interesting region $0.4 < \eta < 1$. This is illustrated in
fig.~\ref{fig:eta-dep}, where the three components of $X_c$ in
eq.~(\ref{Xc}) are plotted as well. The second component is
numerically small, while there is a strong cancellation of the
$\eta$-dependence between the first and the third component. However,
these components are not separately physical in any conceivable limit,
so the cancellation cannot be considered accidental. 

Since $X_c$ is practically scale-independent, $X_t$ must be the source
of the factor of $\;\sim\!\! 3$ enhancement of BR$_\gamma$ by QCD
logarithms. This is indeed the case, because all the powers of $\eta$
in eq.~(\ref{Xt}) are positive and quite large. When $\eta$ changes
from unity to 0.566 (which corresponds to $\mu_0=M_W$ and
$\mu_b=5$~GeV), then $X_t$ decreases from 0.450 to 0.325. At the same
time, $X_c$ changes by only 0.008 (from $-\f{23}{36} \approx -0.639$
to $-0.631$). Consequently, $|C^{(0)\rm eff}_7(\mu)|^2$ increases from
0.036 to 0.094, i.e.\ the branching ratio gets enhanced by a factor of
2.6.

It is easy to identify the reason for the strong $\eta$-dependence of
$X_t$. It is the large anomalous dimension of $m_b(\mu)$ that stands
in front of the operator $P_7$ (\ref{ops}). The anomalous dimension
$\gamma_m$ is responsible for $\f{12}{23}$ out of $\f{16}{23}$ in the
power of $\eta$ that multiplies the (numerically dominant) function
$A_0^t(x)$ in the expression for $X_t$.  Thus, the logarithmic QCD
effects in $b \to s \gamma$ can be approximately taken into account by
simply keeping $m_b$ renormalized at 
$\mu_0 \sim (m_t$~or~$M_W)$ 
in the top contribution to the decay amplitude. In section
\ref{sec:num}, we shall see that those features carry on at the NLO,
lending support to the idea that they are not due to a numerical
coincidence, and remain valid at the NNLO, too.

It would be interesting to understand the physics behind the different
optimal normalizations of $m_b$ in the charm and top loops. In this
respect, the following observations may be helpful.  Off--shell $b\to
s \gamma$ and $b\to s\;gluon$ diagrams mediated by charm loops give
rise only to dimension-six operators
$P_{\sigma F D}=\f{ie}{16\pi^2} \bar{s}_L \{\sigma_{\mu\nu} F^{\nu\mu}, \slash D \} b_L$
~and~
$P_{\sigma G D}=\f{ig}{16\pi^2} \bar{s}_L \{\sigma_{\mu\nu} G^{a\;\nu\mu} T^a, \slash D \} b_L$,
respectively. The basis of physical operators in the off-shell
effective theory for the charm sector can be chosen to be $ \{ P_1,
..., P_6, P_{\sigma F D}, P_{\sigma G D} \} $. The LO Wilson
coefficient of $P_{\sigma F D}$ is given by eq.~(\ref{Xc}), i.e. its
RGE evolution is very slow. When the evolution is terminated at
$\mu_b$, and the equations of motion (EOM) are used afterwards,
$P_{\sigma F D}$ reduces to the operator $P_7$ that contains an
overall factor of $m_b$. Thus, the $b$ mass naturally associated with
charm loops is a low energy, low virtuality mass.
\begin{figure}[htb]
\begin{center}
\includegraphics[width=14cm,angle=0]{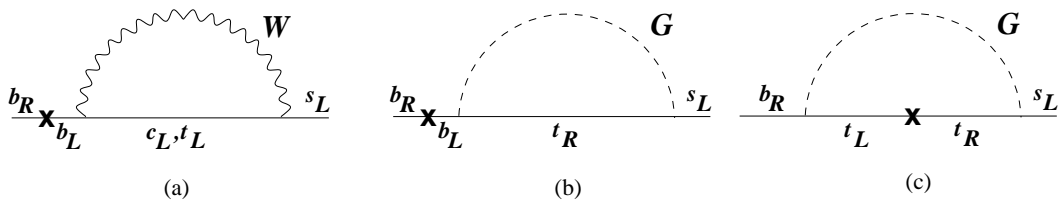}
\caption{\sf Chirality flows in diagrams contributing to $C_7^{(0)}(\mu_0)$.
The photon couples to any of the internal lines.}
\label{fig:PG}
\end{center}
\end{figure}

\ \\[-18mm]

In the top loops, on the other hand, $m_b$ appears in two ways: either
through the same mechanism as in the charm loops
(fig.~\ref{fig:PG}a,b), or via the bottom Yukawa coupling $y_b$ of a
right-handed $b$ with a left-handed top quark (fig.~\ref{fig:PG}c).
In the first case, the RGE evolution of $C^t_{\sigma F D}(\mu)$ and
$C^t_{\sigma G D}(\mu)$ is much faster than in the charm sector. Such
a fast running can be compensated to a large extent by using
$m_b(\mu_0)$ in the EOM at the low--energy scale. In the second case,
it is not necessary to use the EOM to project on $P_7$. The natural
normalization scale of the Yukawa coupling $y_b(\mu)$ is given by the
heavy masses circulating in the loop. The $b$ mass associated with
the Yukawa contributions is therefore a high--energy mass,
$m_b(\mu_0)$ with $\mu_0 \sim (M_W~{\rm or}~m_t)$.  This reasoning
leads us to the conclusion that the appropriate normalization of the
total top contribution is in terms of $m_b(\mu_0)$.

A careful reader might worry about how this picture can be reconciled
with GIM cancellations that take place in the case of degenerate
quarks. Some light on this issue can be shed by considering a
hypothetical situation when $m_b \ll m_c \ll m_t \ll M_W$. In such a
case, the effective $bs\gamma$ interaction would be relatively well
approximated by
\be \label{hypot}
{\cal L}_{\rm int} \approx \f{4 G_F}{\sqrt{2}} V^*_{ts} V_{tb}
\left\{ \f{23}{36} [ m_b(m_t) - m_b(m_c) ] + {\cal O}(m^2_{c,t}/M_W^2) \right\}
\bar{s}_L \sigma^{\mu\nu} b_R F_{\mu\nu}.
\ee
It is not an exact formula even at LO, because we have neglected the
very slow RGE running from $M_W$ to $m_{c,t}$ and the small anomalous
dimensions of $P_{7,8}/m_b(\mu)$. However, eq.~(\ref{hypot}) works
numerically quite well. The QCD enhancement of BR$_\gamma$ can be much
larger here (than in the real world), and its main reason is clearly
seen.  The number $\f{23}{36}$ arises from the matching condition with
an effectively massless internal quark. In the absence of QCD, it
drops out owing to GIM cancellation. However, no cancellation takes place
any longer when the QCD evolution of $m_b$ is taken into account.

\newsection{Next-to-leading order formulae}
\label{sec:NLO}

We have seen in the previous section that most of the leading QCD
logarithms can be taken into account by keeping the top and charm
contributions split, and by renormalizing $m_b(\mu)$ in $P_7$
(\ref{ops}) at $\mu_0 \sim (m_t$~or~$M_W)$ in the top contribution to
the decay amplitude. Motivated by this observation, we shall now
rewrite the known NLO expressions for $\bar{B} \to X_s \gamma$ in an
analogous manner. As we shall see, this simple operation not only
allows us to reproduce the logarithmic QCD enhancement in a
satisfactory manner, but also the NLO corrections become significantly
smaller than in the traditional approach. Moreover, the residual
renormalization-scale-dependence diminishes, without any accidental
cancellations involved. In other words, the behaviour of QCD
perturbation series improves.

Assuming that the dominant NNLO QCD effects have the same origin as
the dominant LO and NLO ones, we shall use all the currently known
perturbative information to determine the ratio of $m_b(\mu_0)$ to the
low--energy $b$-quark mass that normalizes the semileptonic decay rate.
This low--energy mass will be chosen in a manner that ensures good
convergence of QCD-perturbation series for the semileptonic decay.

Our input here are the standard NLO QCD formulae for $B \to X_s
\gamma$ collected in ref.~\cite{CMM97}, and the separate charm-sector
and top-sector matching conditions for the relevant operators
presented in section 2 of ref.~\cite{BMU00.1}. The evaluation of such
separate matching conditions was described in great detail in
section~5 of the latter paper. We have also checked it independently
with the method of ref.~\cite{CDGG98}. Apart from the perturbative QCD
effects, we shall include the electroweak and the available
non-perturbative corrections.

The $\bar{B} \to X_s \gamma$ branching ratio with an energy cut--off
$E_0$ in the $\bar{B}$-meson rest frame can be expressed as follows:
\mathindent0cm
\be \label{main}
{\rm BR}[\bar{B} \to X_s \gamma]^{{\rm subtracted~} \psi,\;\psi'
}_{E_{\gamma} > E_0}
= {\rm BR}[\bar{B} \to X_c e \bar{\nu}]_{\rm exp} 
\left| \f{ V^*_{ts} V_{tb}}{V_{cb}} \right|^2 
\f{6 \alpha_{\rm em}}{\pi\;C} 
\left[ P(E_0) + N(E_0) \right],
\ee
\mathindent1cm
where $\alpha_{\rm em} = \alpha_{\rm em}^{\rm on~shell}$ \cite{CM98}
and $P(E_0)$ is given by the perturbative ratio
\be \label{pert.ratio}
\f{\Gamma[ b \to X_s \gamma]_{E_{\gamma} > E_0}}{
|V_{cb}/V_{ub}|^2 \; \Gamma[ b \to X_u e \bar{\nu}]} = 
\left| \f{ V^*_{ts} V_{tb}}{V_{cb}} \right|^2 
\f{6 \alpha_{\rm em}}{\pi} \; P(E_0).
\ee
$N(E_0)$ denotes the non-perturbative correction.\footnote{
This means that $P(E_0)$ gets replaced by $P(E_0)+N(E_0)$
when $b$ is replaced by $\bar{B}$ in eq.~(\ref{pert.ratio}).}
Contrary to the standard approach, we have chosen the {\em charmless}
semileptonic rate (corrected for the appropriate CKM angles) to be the
normalization factor in eq.~(\ref{pert.ratio}). This modification is
offset by the factor $C$ in eq.~(\ref{main}):
\be \label{phase1}
C = \left| \f{V_{ub}}{V_{cb}} \right|^2 
\f{\Gamma[\bar{B} \to X_c e \bar{\nu}]}{\Gamma[\bar{B} \to X_u e \bar{\nu}]}.
\ee
This observable can either be measured or calculated. Our normalization
to the {\em charmless} semileptonic rate in the l.h.s. of
eq.~(\ref{pert.ratio}) is motivated by the need for separating the
problem of $m_c$ determination from the problem of convergence of
perturbation series in $b \to X_s \gamma$.  The factor $C$ can
be called ``the non-perturbative semileptonic phase-space factor''.

The superscript ``subtracted $\psi$ and $\psi'$'' on the l.h.s. of
eq.~(\ref{main}) means that the processes
\bea
\bar{B} &\to& \psi^( {}' {}^{)} X^{(1)} \nonumber \\[-3.5mm]
&& \hspace{3mm} \searrow \nonumber\\[-5mm]
&& \hspace{7mm} X^{(2)} \gamma\\[-9mm] \nonumber 
\eea
are treated as background, and should be subtracted on the
experimental side. This background is negligible (below 1\%) for a
cut--off of 2.1~GeV on the photon energy in the $\bar{B}$-meson rest
frame, but may have a $\sim 5\%$ effect when the cut--off is lowered to 
1.8~GeV.\footnote{
The effect would become more than 100\% for $E_0 \sim 0.2$~GeV.}

The perturbative quantity $P(E_0)$ can be written in the following form:
\be \label{Pdel}
P(E_0) = \left| K_c + 
\left( 1 + \f{\al(\mu_0)}{\pi} \ln \f{\mu_0^2}{m_t^2} \right)
r(\mu_0) K_t + \varepsilon_{\rm ew} \right|^2 + B(E_0),
\ee
where $K_t$ contains the top contributions to the $b \to s \gamma$
amplitude. $K_c$ contains the remaining contributions, among which the
charm loops are by far dominant. The electroweak correction to the
$b \to s \gamma$ amplitude is denoted by $\varepsilon_{\rm ew}$. The
ratio
\be \label{mratio} 
r(\mu_0) = \f{m_b^{\overline{\rm MS}}(\mu_0)}{m_b^{1S}} 
\ee
appears in eq.~(\ref{Pdel}) because we keep $m_b$ renormalized at
$\mu_0$ in the top contribution to the operator $P_7$ (\ref{ops}),
while all the kinematical factors of $m_b$ are expressed in terms of
the bottom ``1S mass''. The 1S mass is defined as half of the
perturbative contribution to the $\Upsilon$ mass. As argued in
ref.~\cite{HLM99}, expressing the kinematical factors of $m_b$ in
inclusive $B$-meson decay rates in terms of the 1S mass improves
the behaviour of QCD perturbation series with respect to what 
would be obtained using $m_b^{\overline{\rm MS}}(m_b)$ or $m_b^{\rm pole}$.

The bremsstrahlung function $B(E_0)$ contains the effects of $b \to s
\gamma g$ and $b \to s \gamma q \bar{q}$ \linebreak $(q=u,d,s)$
transitions. It is the only $E_0$-dependent part in $P(E_0)$.  It is
given in appendix~E. Its influence on the $b \to X_s \gamma$ branching
ratio is less than 4\% when $1~{\rm GeV} < E_0 < 2~{\rm GeV}$.
Therefore, we do not split the top and charm contributions to this
function.  It would not improve the overall accuracy at all, but only
make the formulae unnecessarily complicated.
\begin{table}[t]
\begin{tabular}{|l|r|r|r|r|r|r|r|r|}
\hline
~~~$k$ & 1 & 2 & 3 & 4 & 5~~~~ & 6~~~~ & 7~~~~ & 8~~~~ \\ 
\hline
\ &&&&&&&& \\[-4mm]
$a_k$              & $\f{14}{23}$~~~  &  $\f{16}{23}$~~~ & $\f{6}{23}$~~~ & $-\f{12}{23}$~~~ 
                   &     0.4086 & $-$0.4230 & $-$0.8994 &    0.1456 \\[1.5mm]
$d_k$              &     1.4107 & $-$0.8380 & $-$0.4286 & $-$0.0714 
                   &  $-$0.6494 & $-$0.0380 & $-$0.0185 & $-$0.0057 \\[1.5mm]
$\tilde{d}_k$      & $-$17.6507 &  11.3460  &    3.5762 & $-$2.2672
                   &     3.9267 &   1.1366  & $-$0.5445 &    0.1653 \\[1.5mm]
$\tilde{d}^{\eta}_k$&    9.2746 & $-$6.9366 & $-$0.8740 &    0.4218
                   &  $-$2.7231 &    0.4083 &    0.1465 &    0.0205 \\[1.5mm]
$\tilde{d}^{a}_k$  &   0~~~~~~~~&  0~~~~~~~~&  1~~~~~~~~&  1~~~~~~~ 
                   &   0~~~~~~~~&  0~~~~~~~~&  0~~~~~~~~&  0~~~~~~~~\\[1.5mm] 
$\tilde{d}^{b}_k$  &   0~~~~~~~~&  0~~~~~~~~&  1~~~~~~~~&  1~~~~~~~ 
                   &   0~~~~~~~~&  0~~~~~~~~&  0~~~~~~~~&  0~~~~~~~~\\[1.5mm] 
$ \tilde{d}^{i \pi}_k$&  0.4702 &  0~~~~~~~~& $-$0.4938 & $-$0.4938
                   &  $-$0.8120 &    0.0776 & $-$0.0507 &    0.0186 \\[1.5mm]
$ e_k$             &     5.2620 & $-$3.8412 &  0~~~~~~~~& 0~~~~~~~
                   &  $-$1.9043 & $-$0.1008 &    0.1216 &    0.0183 \\[1.5mm]
\hline
\end{tabular}
\caption{\sf ``Magic numbers'' for $K_c$ and $K_t$}
\label{tab:mag}
\end{table}

\ \\[-1cm]
Our result for $K_c$ reads
\bea 
K_c &=& \sum_{k=1}^8 \eta^{a_k} \left\{ d_k + \f{\al(\mu_b)}{4\pi} \left[
2 \beta_0 a_k d_k \left( \ln \f{m_b}{\mu_b} + \eta \ln \f{\mu_0}{M_W} \right)
\right. \right. \nonumber\\[2mm] && \hspace{-11mm} \left. \left. 
+ \tilde{d}_k + \tilde{d}^{\eta}_k \; \eta
+ \tilde{d}^{a}_k a(z)  + \tilde{d}^{b}_k b(z)  
+ \tilde{d}^{i \pi}_k i \pi \right] \right\} 
\;\;+ \f{V_{us}^* V_{ub}}{V_{ts}^* V_{tb}} \, \f{\al}{4\pi} \,
\left( \eta^{a_3} + \eta^{a_4} \right)  [ a(z) + b(z) ],
\label{Kc}
\eea
where $\beta_0 = \f{23}{3}$ and $z = (m_c/m_b)^2$.
The functions $a(z)$ and $b(z)$ are given in appendix D.

The ``magic numbers'' $d_k,~\tilde{d}_k,~...$ can be found in
table~\ref{tab:mag}. In their evaluation, the unknown NLO matrix
elements $r_3, ..., r_6$ \cite{CMM97} have been set to zero. The
resulting error will be absorbed in the NNLO uncertainty below. This
is mandatory\footnote{
We have checked that those parts of $r_3, ..., r_6$ that are not due
to closed $b$-quark loops have only a 0.5\% effect on BR$_\gamma$.}
because ~~$C^{(0)}_k(\mu_b)/C^{(0)}_2(\mu_b) \leq
\f{\al(\mu_b)}{\pi}$~ for ~$k=3, ..., 6$.

The NLO expression for $K_t$ is as follows
\mathindent0cm
\bea
K_t &=& \left[ 1 -\f{2}{9} \al(m_b)^2 +\f{\al(\mu_0)}{\pi}
\ln \f{\mu_0}{m_t} \;\; 4 x \f{\partial}{\partial x} \right] 
\left[ -\f{1}{2} \eta^{\f{4}{23}} A^t_0(x) 
+ \f{4}{3} \left( \eta^{\f{4}{23}} - \eta^{\f{2}{23}} \right) F^t_0(x) \right] 
\nonumber\\&& \hspace{-15mm}
+ \f{\al(\mu_b)}{4\pi} \left\{ E_0^t(x) \sum_{k=1}^8 e_k \eta^{\left(a_k+\f{11}{23}\right)}
\right. \nonumber\\ && \left. \hspace{-5mm}
+ \eta^{\f{4}{23}} \left[ -\f{1}{2} \eta A^t_1(x) 
+ \left( \f{12523}{3174} ~~-\f{7411}{4761} \eta ~~~~-\f{2}{9} \pi^2
      ~~-\f{4}{3} ~ \left( \ln \f{m_b}{\mu_b} + \eta \ln \f{\mu_0}{m_t} \right) \right) A^t_0(x) 
\right. \right. \nonumber\\ && \left. \left. \hspace{6mm}
                           +\f{4}{3} \eta F^t_1(x)
+ \left( -\f{50092}{4761} +\f{1110842}{357075} \eta +\f{16}{27} \pi^2
         +\f{32}{9} \left( \ln \f{m_b}{\mu_b} + \eta \ln \f{\mu_0}{m_t} \right) \right) F^t_0(x) 
\right]
\right. \nonumber\\ && \left. \hspace{-5mm}
+ \eta^{\f{2}{23}} \left[ -\f{4}{3} \eta F^t_1(x)
+ \left( \f{2745458}{357075} -\f{38890}{14283}  \eta -\f{4}{9} \pi (\pi+i)
         -\f{16}{9} \left( \ln \f{m_b}{\mu_b} + \eta \ln \f{\mu_0}{m_t} \right) \right) F^t_0(x) 
\right] \right\}. \nonumber\\ 
\label{Kt}
\eea
\mathindent1cm
The functions $A^t_0$ and $F^t_0$ of $x=(m_t(\mu_0)/M_W)^2$ have
already been given in eq.~(\ref{A0.andF0}). The remaining functions read
\bea \begin{array}{rcl}
E^t_0(x) &=& \f{-9x^2+16x-4}{6(x-1)^4} \ln x  
+\f{7x^3+21x^2-42x-4}{36(x-1)^3},\\[2mm]
A^t_1(x) &=& \f{32x^4+244x^3-160x^2+16x}{9(1-x)^4} Li_2\left(1-\f{1}{x}\right) 
+ \f{-774x^4-2826x^3+1994x^2-130x+8}{81(1-x)^5} \ln x
\\[4mm] &&
+ \f{-94x^4-18665x^3+20682x^2-9113x+2006}{243(1-x)^4},\\[2mm]
F^t_1(x) &=& \f{4x^4-40x^3-41x^2-x}{3(1-x)^4} Li_2\left(1-\f{1}{x}\right) 
+\f{-144x^4+3177x^3+3661x^2+250x-32}{108(1-x)^5} \ln x
\\[4mm] &&
+\f{-247x^4+11890x^3+31779x^2-2966x+1016}{648(1-x)^4}.
\end{array} \nonumber
\eea     
The ${\cal O}(\al^2)$ term is included in $K_t$ along the lines of the
 $\Upsilon$ expansion \cite{HLM99}. Its effect is at the level of 1\%
  only, and it is offset by an analogous term in $r$ (\ref{mratio})
(see appendix B).

The electroweak correction $\varepsilon_{\rm ew}$ in eq.~(\ref{Pdel})
consists of three terms
\be \label{ewcor}
\varepsilon_{\rm ew} = \delta^{\rm ew} C_7^{(0)\rm eff}(\mu_b)
    + \f{\alpha_{em}(M_Z)}{\al(\mu_b)} C_7^{\rm em(0)eff}(\mu_b)
    - \f{\alpha_{em}(M_Z)}{\pi} \left( K_c^{(0)} + r K_t^{(0)} \right)\ln \f{M_Z}{\mu_b}.
\ee
The first term stands for the dominant non--logarithmic electroweak
effects calculated in ref.~\cite{GH00} (see eq.~(11) of that paper).
The remaining two terms originate from the logarithmically enhanced
electromagnetic corrections to the weak radiative (eq.~(12) of
ref.~\cite{BM00}) and  semileptonic (eq.~(1) of ref.~\cite{S81})
decays, respectively.  In the last term, $K_c^{(0)}$ and $K_t^{(0)}$
are just the LO contributions to $K_c$ and $K_t$.

The non-perturbative correction $N(E_0)$ in eq.~(\ref{main}) is
partly known
\be \label{ndel}
N(E_0) = -\f{1}{18} \left( K_c^{(0)} + r K_t^{(0)} \right)
\left( \eta^{\f{6}{23}} + \eta^{-\f{12}{23}} \right) \; \f{\lambda_2}{m_c^2} 
+ ...
\ee
The calculable $\Lambda^2/m_c^2$ correction \cite{1.ov.mc} is taken
into account above, while the calculable $\Lambda^2/m_b^2$ corrections
\cite{1.ov.mb} have cancelled out because of our normalization to the
{\em charmless} semileptonic rate. The dots stand for higher-order
terms in the heavy-quark expansion \cite{1.ov.mc,B98}, as well as for
the non-perturbative effects due to higher (than $\psi$ and
$\psi'$) intermediate $\bar{c}c$ states, to light-quark loops and to
motion of the $b$-quark inside the $\bar{B}$ meson. In our numerical
results (calculated with $E_0=1.6$~GeV), we shall set those effects to
zero without including any additional uncertainty.

\vfill

\newsection{Numerical analysis}
\label{sec:num}
 
In the present section, we shall test our formulae numerically.  The
experimental inputs and the ranges for the renormalization scales are
collected in appendix A. Appendices B, C and D are devoted to the
determination of the mass ratio $r$ (\ref{mratio}), the phase-space
factor C (\ref{phase1}) and the $z$-dependent terms in $K_c$
(\ref{Kc}), respectively. The final results obtained there read
\bea
r(\mu_0 = m_t) &=& 0.578 \pm 0.002_{\mu_b} \pm (\mbox{parametric errors}) \\
C &=& 0.575 \; ( 1 \pm 0.01 \pm 0.02 \pm  0.02 )\\
a(z) &=& (0.97 \pm 0.25) \;+\; i ( 1.01 \pm 0.15 )\\
b(z) &=& (-0.04 \pm 0.01) \;+\; i ( 0.09 \pm 0.02 ).
\eea
In our NLO computation, the imaginary parts of $K_c$ and $K_t$ are
irrelevant, because all the ${\cal O}(\al^2)$ terms on the r.h.s. of
eq.~(\ref{Pdel}) are set to zero after the square is taken. If the
imaginary parts of $K_c$ and $K_t$ were not set to zero, they would
affect $P(E_0)$ by only 0.5\%.

The ratio of CKM angles standing in front of $P(E_0)$ can be
expressed in terms of the Wolfenstein parameters as follows
\bea
\left| \f{ V^*_{ts} V_{tb}}{V_{cb}} \right|^2 &=& 
1 + \lambda^2 (2 \bar{\rho}-1) + \lambda^4 (\bar{\rho}^2+\bar{\eta}^2-A^2)
+ {\cal O}(\lambda^6) \nonumber\\
&\approx& 0.971 + 0.10 (\bar{\rho}-0.224) = 0.971 \pm 0.004,
\label{CKMratio}
\eea
where we have used $\lambda = 0.2237$, $A=0.819$, $\bar{\rho}=0.224
\pm 0.038$ and $\bar{\eta} = 0.317$~ \cite{CAFLMPRS00}. The only
relevant source of uncertainty is the error in ${\bar{\rho}}$. Even if
this error were enlarged by a factor of 2, the influence of $|V^*_{ts}
V_{tb}/V_{cb} |^2$ on the overall uncertainty in $\bar{B} \to X_s
\gamma$ would remain negligible. The central value of
eq.~(\ref{CKMratio}) is also consistent with the analysis of
ref.~\cite{mele}.

The electroweak and non-perturbative corrections from
eqs.~(\ref{ewcor}) and (\ref{ndel}) take the following values\footnote{
  The electroweak correction remains practically unchanged
  ($\varepsilon_{\rm ew} = 0.0071$) when the very recent results of
  ref.~\cite{GH01} are included.}
\bea 
\varepsilon_{\rm ew} &\approx& 0.0035 + 0.0012 + 0.0028 = 0.0075 \label{ewcor.num} \\
N(E_0) &=& 0.0036 \pm 0.0006 \,.   \label{ndel.num}
\eea
Their effects on BR$_\gamma$ are $-3.8\%$ and $+2.5\%$, respectively.
In the electroweak correction, $M_{\rm Higgs} = 115$~GeV has been
used. The sensitivity of our final result (\ref{main.num}) to $M_{\rm
Higgs}$ is very weak (only 0.3\% when $M_{\rm Higgs}$ changes from
115~GeV to 200~GeV). In $N(E_0)$, $\lambda_2 = 0.12~{\rm GeV}^2$ and
$m_c = m_c(m_c) = (1.25 \pm 0.10)$~GeV \cite{PData00} have been
used. The indicated uncertainty is due to $m_c$ only.

In the ``naive'' approach, when only the one--loop electroweak diagrams
are calculated, one finds
\be 
K_c^{\rm naive} = -\f{23}{36} \approx -0.639
\hspace{1cm} {\rm and} \hspace{1cm}
K_t^{\rm naive} = -\f{1}{2} A^t_0(x) \approx 0.450.
\ee
From those numbers, we obtain
\be
K_c^{\rm naive} + \f{m_b(m_t)}{m_b^{1S}} K_t^{\rm naive} \approx -0.379,
\ee
which implies BR$^{\rm naive}_{\gamma} \approx 3.53 \times 10^{-4}$
~(for $N(E_0)=0$). If the factor $m_b(m_t)/m_b^{1S}$ were not
included, our ``naive'' prediction would be 4 times lower.\footnote{
If ~$m_b(M_W)/m_b(5~{\rm GeV}) \approx 0.713$~ were used instead of
~$m_b(m_t)/m_b^{1S} \approx 0.578$~ in $P(E_0)$ (\ref{Pdel}), we would
obtain ~BR$^{\rm naive}_{\gamma} \approx 2.49 \times 10^{-4}$.}
 
At LO, i.e. when the ${\cal O}(\al)$ terms in eqs.~(\ref{Kc}) and
(\ref{Kt}) are neglected, we find
\be
K_c^{(0)} = -0.631^{-0.003}_{+0.000} 
\hspace{1cm} {\rm and} \hspace{1cm}
K_t^{(0)} = 0.434^{-0.005}_{+0.004},
\ee
which implies that BR$_{\gamma}^{\rm LO} = 3.56^{+0.14}_{-0.07} \times
10^{-4}$. The indicated errors correspond to the variation of the
low-energy scale $\mu_b$, as described in appendix A.

At NLO, we find for $E_0 = 1.6$~GeV
\bea 
K_c &=& -0.611^{+0.002}_{-0.001} +i\; \left(-0.032^{-0.009}_{+0.006} \right),\label{numkc}\\  
K_t &=& \hspace{3mm} 
         (0.397 \pm 0.003) + i\; ( 0.011 \pm 0.002), \label{numkt}\\ 
B(E_0) &=& 3.1^{+1.3}_{-0.7} \times 10^{-3}.
\eea
From the above three results, we obtain
BR$[\bar{B} \to X_s \gamma]_{E_{\gamma} > 1.6~{\rm GeV}} 
= 3.60^{+0.04}_{-0.05} \times 10^{-4}$.  
As before, the quoted errors are due to $\mu_b$ only.

We can see that the $\mu_b$-dependence of ${\rm Re}(K_c)$ and ${\rm
Re}(K_t)$ is very weak, both at LO and at NLO. Such a weak
scale-dependence is somewhat surprising at LO, but at NLO it is
not. Moreover, at NLO, the weak scale-dependence is not a result of
any accidental cancellation among strongly $\mu_b$-dependent
terms. Whatever cancellations occur, they occur inside the charm
contribution $K_c$, and thus cannot be considered accidental.

Since the $\mu_b$-dependence of the NLO branching ratio is very weak
(below $\pm 1.4\%$), and the dependence on $\mu_0$ is not stronger, we
can estimate the NNLO corrections by simply saying that they are of
order $(\al(m_b)/\pi)^2 \approx 0.5\%$, up to an unknown factor of
order unity. Thus, it seems safe to assume a value of $\pm 4\%$ for
the theoretical error which is due to neglecting $r_3$, ..., $r_6$ at
NLO and to the NNLO effects.  This is almost twice the combined
scale-dependence of the result obtained by scanning $\mu_b$ and
$\mu_0$ between one half and twice their central values (2.2\%).  We
consider the error related to the value of $m_c/m_b$ in $K_c$
separately. The latter uncertainty amounts to $\pm 5.5\%$ alone, when
our estimate from appendix D is used.

When all the sources of uncertainties are included, we find
\bea 
{\rm BR}[\bar{B} \to X_s \gamma]^{{\rm subtracted~} \psi,\;\psi'
}_{E_{\gamma} > 1.6~{\rm GeV}}
&=& 3.60 \times 10^{-4} \; ( 1 
\pm 0.06_{(m_c/m_b~{\rm in}~K_c)} 
\pm 0.04_{\rm (other~NNLO)}
\nonumber \\ && \hspace{27mm}
\pm 0.01_{\rm (pert~C)}
\pm 0.02_{\rm \lambda_1}
\pm 0.02_{\rm \Delta} 
\nonumber \\ && \hspace{27mm}
\pm 0.02_{\al(M_Z)} 
\pm 0.02_{{\rm BR}({\rm semilept})_{\rm exp}}
\pm 0.01_{m_t})
\nonumber \\
&=& (3.60 \pm 0.30) \times 10^{-4}.
\label{main.num}
\eea
The errors in the second line above originate from the semileptonic
phase-space factor in eq.~(\ref{phase.num}). The last of them is due
to the error in $m_b^{1S}$. Other effects caused by the uncertainty in
$m_b^{1S}$ are negligible (below 0.2\%).

In the last line of eq.~(\ref{main.num}), all the uncertainties have
been added in squares. Thus, the final error has only an illustrative
character, because many of the partial ones have no statistical
interpretation.

We have chosen $E_0=1.6$~GeV instead of the commonly used $E_0 =
\f{1}{20} m_b \approx 0.23$~GeV (i.e. $\delta \equiv 1 - 2 E_0/m_b =
0.9$), because the low-energy photons in $\bar{B} \to X_s \gamma$ are
not experimentally accessible.  For high values of the cut--off (around
and above 2 GeV), non-perturbative uncertainties are much larger
because the $\bar{B}$--meson shape functions are unknown. Models for
the shape functions considered in refs.~\cite{KN99,BLM01} suggest that
$E_0=1.6$~GeV is already low enough to make those uncertainties
negligible. In our opinion, the way to resolve this issue is to measure
the photon spectrum without any theoretical input, with the photon
energy cut--off as low as possible (e.g. 2.0 or 1.9 GeV in the 
$\bar{B}$--meson rest frame). Next, a simple extrapolation to the theoretically
known spectrum in the vicinity of 1.6--1.7 GeV should be made. For
this purpose, the following approximate expression for the integrated
branching ratio as a function of $E_0$ might be useful:
\be
BR(E_0) = (3.524 + 0.298 E_0 - 0.157 E_0^2 ) \times 10^{-4},
\ee
where $E_0$ is expressed in GeV. The above expression has been obtained
as a fit to our results in the region 1~GeV~$ < E_0 < $~2~GeV. The
accuracy of this fit is $\pm 0.3\%$. Outside the considered region of
$E_0$, our formulae from section \ref{sec:NLO} are not expected to
work well. On the low--energy side, it is due to the fact that we have
not included the fragmentation functions discussed in
ref.~\cite{KLP95}. As far as the high--energy side is concerned, the
growth of uncertainties with $E_0$ can be roughly estimated by using
fig.~3 of ref.~\cite{KN99} and fig.~3 of ref.~\cite{BLM01}. However,
translating those plots into quantitative estimates is rather
difficult, because of the model-dependence involved. A further study
of this issue is crucial for a precise comparison of theory and
experiment in $\bar{B} \to X_s \gamma$.

In view of the fact that many of the published results have been
calculated for $\delta=0.9$ (i.e. $E_0 = \f{1}{20} m_b$), it is
interesting to check what our formulae give in such a case. We find
\be
B\left( \mbox{$\f{1}{20}$} m_b \right) = 8.2^{+3.7}_{-2.2} \times 10^{-3}.
\ee
Together with eqs.~(\ref{numkc}) and (\ref{numkt}), this gives
\be \label{br.tot}
BR[\bar{B} \to X_s \gamma]_{E_{\gamma} > m_b/20} 
= 3.73 \times 10^{-4}.  
\ee
The total relative error in this case is roughly comparable to the one
given in eq.~(\ref{main.num}). If we used $m_c/m_b = 0.29$ instead of
0.22 in $K_c$ and $B(E_0)$, we would find $3.35 \times 10^{-4}$ for
the branching ratio.  The latter result is very close to the ones
obtained in many previous analyses (see e.g. \cite{KN99,GH00}). Thus,
the replacement of $m_c^{\rm pole}/m_b^{\rm pole}$ by
$m_c^{\overline{\rm MS}}(\mu)/m_b^{1S}$ in $\me{P_2}$
is the main reason why our result is significantly higher than the
previously published ones.

Our NLO formulae differ from the ones used previously  by residual
${\cal O}(\al^2)$ terms. Quite unexpectedly, those terms tend to
cancel among themselves, even though they are not individually very
small. For instance, if we used eq.~(\ref{mratio.NLO}) instead of
(\ref{mratio.num}) for the determination of the ratio $r$, our final
result for the branching ratio would be 7\% lower than the one in
eq.~(\ref{main.num}).

\newsection{Constraints on new physics}
\label{sec:newphys}

In the absence of new light degrees of freedom, physics beyond the SM
manifests itself through (i) new contributions to coefficients of the
operators involved in the SM calculation and \linebreak (ii) the
appearance of operators absent in the SM, such as operators with
different chirality \cite{borz2}.  Examples of contributions beyond
the SM include diagrams like the ones in fig.~1, with charged Higgs
exchange or with chargino--squark loops.  Therefore, new physics
contributions are at the same level as the SM ones, and similar
theoretical accuracy is necessary.

In all the new physics scenarios which do not involve new operators
(and in which \linebreak $C_k^{\rm new}(\mu_0) = 0$~ for
~$k=1,2,3,5,6$), it is straightforward to incorporate extra
contributions to the NLO formulae of section \ref{sec:NLO}.  These
contributions effectively modify eqs.~(\ref{Kt})\footnote{
Except for the terms proportional to ~$\ln(\mu_0/m_t)$~ in eq.~(\ref{Kt}).
Such terms should be left unchanged, because the appropriate
logarithms of $\mu_0$ are already contained in
$C_{7,8}^{(1)\rm new}(\mu_0)$.}
and (E.9) through the replacements
\bea
A_j^t(x) & \rightarrow  &A_j^t(x) - 2 \,C_7^{(j)\,\rm new}(\mu_0),
\hspace{.7cm}  
F_j^t(x)\   \rightarrow  \ F_j^t(x) - 2 \,C_8^{(j)\,\rm new}(\mu_0),
\hspace{.9cm}    
(j=0,1)\nonumber\\
E_0^t(x)  & \rightarrow & E_0^t(x) +   C_4^{(1)\,\rm new}(\mu_0),
\label{shifts}
\eea
where $C_i^{(j)\rm new}$ are the LO ($j=0$) and NLO ($j=1$) new physics
contributions to the Wilson coefficients of the operators $P_i$
\be
C_i^{\rm new}(\mu_0)=C_i^{(0)\rm new}(\mu_0)+ \frac{\alpha_s(\mu_0)}{4\pi}
\,C_i^{(1)\rm new}(\mu_0).
\ee
Beyond the SM, the necessary NLO coefficients are known at present in
 the general Two--Higgs--Doublet--Model (2HDM)
 \cite{CDGG98,borz,ciaf}, in some specific {SUSY} scenarios
 \cite{CDGG98,DGG00,BMU00.2,Carena:2001uj}, and in the left-right symmetric models
 \cite{BMU00.2}.  The coefficients in (\ref{shifts}) are evaluated at
 the scale $\mu_0\sim M_W, m_t$.  

Whenever the new particles have masses of order $\Lambda\gg
 M_W$, they should be decoupled at $\mu\sim\Lambda$. The coefficients
 $C_i^{(j)\,\rm new}$ should be then evolved down to the scale $\mu_0$, in
 order to resum all large logarithms correctly. Examples of this
 procedure can be found in refs.~\cite{DGG00,anlauf}.

Notice that the coefficients in eq.~(\ref{shifts}) are {\it not}
  the effective ones \cite{BMMP94}, in terms of which the matching
  conditions are often given.  The relation between the two sets is very
  simple:
\be
C_{i}^{\rm eff}(\mu)= C_{i}(\mu) + \sum_{j=1}^6 y_{ij} C_j(\mu),
\ee
where $y_{7j}=(0,0,-\frac13,-\frac49,-\frac{20}3,-\frac{80}{9})$,
$y_{8j}=(0,0,1,-\frac16,20,-\frac{10}{3})$, and all the other $y_{ij}$
vanish.
\newcommand{\gsim}{\;\rlap{\lower 3.5 pt \hbox{$\mathchar \sim$}} \raise 1pt
 \hbox {$>$}\;}
\begin{figure}[h]
\vspace{.4cm}
\begin{center}
\includegraphics[width=11cm,angle=0]{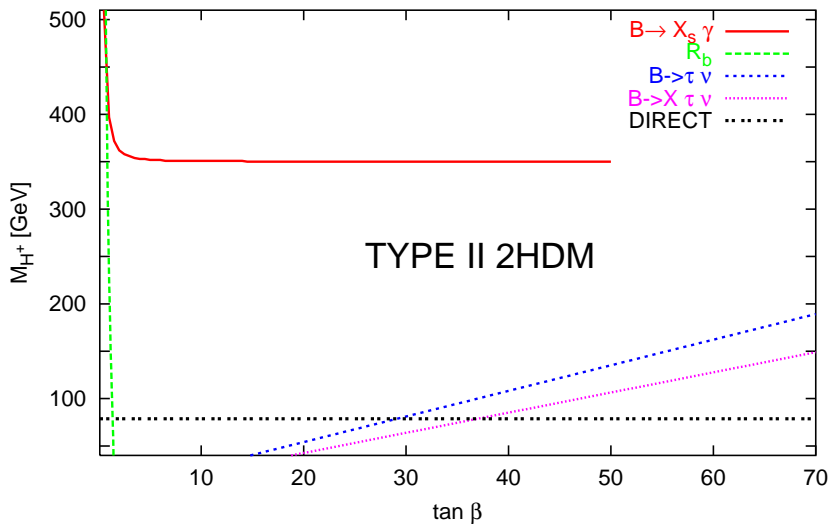}
\end{center}
\vspace{-.7cm}
\caption{\sf Direct and indirect lower bounds on $M_{H^+}$ from different processes in
type II 2HDM as a function of $\tan\beta$. The $B\to X_s \gamma$ bound
is the one in eq.~(\ref{hbound}) below.
\label{fig:hbounds}}
\end{figure}

To illustrate the incorporation of new physics in our calculation of
$\bar{B} \to X_s \gamma$, we consider the case of the 2HDM, which also
exemplifies well the importance of this decay mode for new physics.  We
shall update the lower bound on the charged Higgs mass in the type II
model.

In the 2HDMs with vanishing tree level FCNCs, the only additional
contribution with respect to the SM comes from the charged Higgs
boson--top loops. It depends on the mass of the charged Higgs boson,
$M_H$, and on the ratio of the v.e.v's of the two Higgs doublets,
$\tan \beta $.  Models of type I and II differ by the way fermions
couple to the Higgs doublets. In the type II model (realized in the
MSSM), the charged Higgs loops always enhance BR$_\gamma$, while the
decoupling occurs slowly. Therefore, this decay mode provides strong
lower bounds on $M_H$, whose dependence on $\tan \beta $ saturates for
$\tan \beta\approx 5$. Previous calculations led to $M_H\gsim
250$~GeV, independently of $\tan \beta$ \cite{CDGG98,durham,borz}.
This bound is much stronger than the one from direct searches at LEP2
($M_H> 78.5$~GeV \cite{Holzner}), and than the indirect lower limits
from a number of other processes (see
fig.~\ref{fig:hbounds})~\cite{taunu,Xtaunu}. In model I, $B\to X_s \gamma$
is less restrictive than other processes, because the charged Higgs
loops tend to suppress the branching ratio and decouple for large
$\tan\beta$. The most important constraint in that case comes from
$R_b$ \cite{CDGG98}.

The LO and NLO Wilson coefficients in the 2HDM are given in
eqs.~(52)--(64) of the first paper in \cite{CDGG98}. Adopting the same
notation, with $y= (m_t(\mu_0)/M_H)^2$ and $\mu_W\equiv\mu_0$, we
perform the following replacements in $K_t$
\bea
A_0^t(x) & \rightarrow  &A_0^t(x) - 2 \,\delta C_7^{(0)\rm eff}(\mu_0),
\hspace{1.5cm}    F_0^t(x)  \ \rightarrow  \ 
F_0^t(x) - 2 \,\delta C_8^{(0)\rm eff}(\mu_0),
 \nonumber   \\
E_0^t(x)  & \rightarrow & E_0^t(x) +   E^H(y),\nonumber\\
A_1^t(x) & \rightarrow  &A_1^t(x) - 2 \,\left[G_7^H(y)
  +\Delta_7^H(y)\ln \frac{\mu_0^2}{M_H^2}\right],\\
F_1^t(x)  & \rightarrow  &F_1^t(x) - 2 \,\left[G_8^H(y)+ \Delta_8^H(y)
 \ln\frac{\mu_0^2}{M_H^2} \right].
\nonumber
\label{shifts2HDM}
\eea
As mentioned above, the $\ln \mu_0$ terms in eq.~(\ref{Kt}) should be
left unchanged.
\begin{figure}[h]
\vspace{.4cm}
\begin{center}
\includegraphics[width=11cm,angle=0]{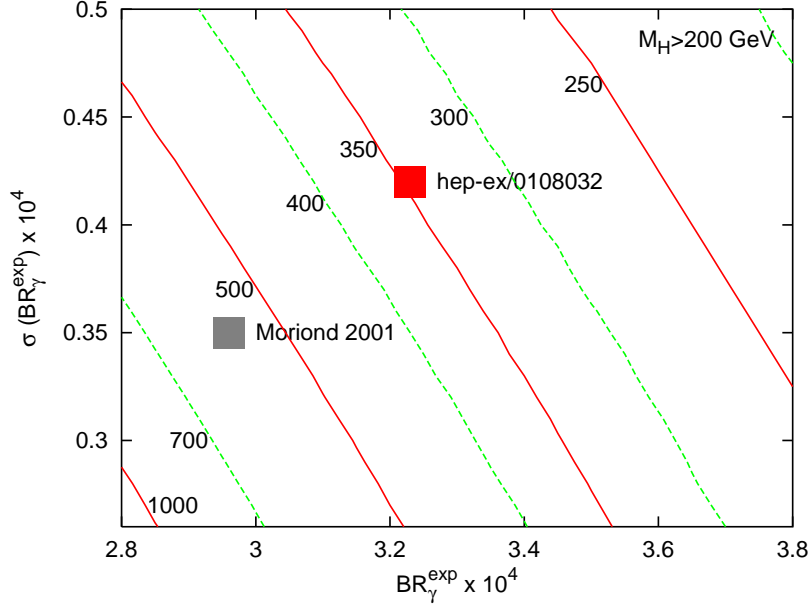}
\end{center}
\vspace{-.7cm}
\caption{\sf The 99\% CL bound on the 2HDM-II charged Higgs mass from
$B\to X_s \gamma$ as a function of the world average and of its
error. The contour lines represent values which lead to the same
$M_{H^+}$ bound. The experimental world averages evaluated with use of
the preliminary \cite{Moriond} and published \cite{CLEO} CLEO results are
indicated for reference.
\label{hcontours}}
\end{figure}

We are now in a position to calculate the branching ratio for $B\to
X_s \gamma$ in the 2HDM, and to find the lower bound on $M_H$ in model
II.  Since the CLEO and BELLE results (obtained with
$E_\gamma>2.0$~GeV and 2.1~GeV, respectively, in the $\Upsilon$ frame)
have been extrapolated to the ``total'' rate, we compare the
experimental result with our prediction for $E_\gamma> \f{1}{20} m_b$
\cite{KN99}.  With respect to previous analyses, the SM prediction is
now higher --- see eq.~(\ref{br.tot}).  Moreover, the charged Higgs
contribution in model II cannot help reducing the value of
BR$_\gamma$. Thus, our bound is going to be stronger than in the past
\cite{CDGG98,durham,borz}.  In addition, one should take into account
that for a heavy charged Higgs, the dependence of BR$_\gamma$ on $M_H$
becomes very mild, signalling the decoupling. Consequently, a small
change in BR$_\gamma$ affects the $M_H$ bound in a significant way.

Therefore, we adopt a very conservative approach.  We employ the
experimental world average $(3.23\pm0.42)\times 10^{-4}$ and scan over
the main theoretical errors in our calculation --- first line of
eq.~(\ref{main.num}) --- notably the one related to $m_c/m_b$ in
$\me{P_2}$.  Whenever the $\mu_b$- and $\mu_0$-dependences of
$BR_\gamma$ in the 2HDM are larger than our 4\% NNLO uncertainty in
eq.~(\ref{main.num}), we expand this error accordingly.\footnote{
Here, the scales are varied between 0.4 and 2.5 times their central values.}
All the remaining parametric and experimental errors are combined in
quadrature.  The absolute lower limit is 
\be \label{hbound}
M_H>350~{\rm GeV}
\ee
for any value of $\tan\beta$.  As far as the gaussian errors are
concerned, this is a 99\% CL bound. The 95\% CL bound is 500~GeV. In
view of future changes in the experimental situation, we display in
fig.~\ref{hcontours} the 99\% CL bounds as functions of the central
value and of the error of the world average.  The dependence of the
$M_H$ lower limit on $\tan\beta$ is shown in fig.~\ref{fig:hbounds}.

The bound is clearly very sensitive to the way various errors are
combined.  For instance, if we combined {\em all} the errors in
quadrature, the 99\% CL absolute lower limit on $M_H$ would be
500~GeV.  On the other hand, the way we combine the errors emphasizes
the large uncertainty coming from $m_c(\mu)/m_b^{1S}$ in $\me{P_2}$.
In effect, the lower limits that we have quoted correspond to
$m_c(m_c)/m_b^{1S}=0.26$, i.e. the highest value of
$m_c(\mu)/m_b^{1S}$ that is compatible with our analysis in appendix
D. Had we employed $m_c/m_b=0.29\pm 0.02$ as in previous analyses and
treated this error as in the derivation of eq.~(\ref{hbound}), the
99\% (95\%) CL bound would have been $M_H > 280~(380)$~GeV.

In deriving the limits on $M_H$, we have not assumed that the 2HDM is
valid for sure, i.e. we have not subtracted 
$\chi^2_{{}_{\rm best\;fit}}$ 
from the $\chi^2$ used for the test. Consequently, the
statement that $M_H > M_0$ at $x\%$~CL means that the 2HDM is excluded
at $x\%$~CL unless $M_H > M_0$. A different procedure\footnote{
We thank A.~Strumia for bringing this point to our attention.}
consists in assuming the validity of the 2HDM and in testing $\chi^2 -
\chi^2_{{}_{\rm best\;fit}}$. In this case, adopting the same
treatment of errors as in eq.~(\ref{hbound}), we obtain 
$M_H > 345~(490)$~GeV at 99\% (95\%) CL. 

\newsection{Conclusions}
\label{sec:concl}

The reanalysis of $\bar{B} \to X_s \gamma$ performed in the present
article contains several major improvements with respect to the
previous ones. In particular, the semileptonic phase--space factor is
expressed in terms of an observable for which the NNLO results are known.
Moreover, the RGE evolution of $m_b(\mu)$ in the top--quark
contribution to the $b \to s \gamma$ amplitude is identified as the main
reason for the huge enhancement of the branching ratio by QCD
logarithms.  In consequence, better control over the behaviour of QCD
perturbation series is achieved. Unfortunately, the desired reduction
of theoretical uncertainties significantly below the current
experimental ones is found to be impossible at the NLO level, because
of the strong $m_c$-dependence of certain two--loop diagrams with
charm-quark loops.

In order to compare our final prediction (\ref{main.num}) with the
weighted average (\ref{main.exp}) of the experimental results, we have
either to evaluate the theoretical errors in eq.~(\ref{br.tot}) or
rather to ``take back'' the extrapolation to low photon energies on
the experimental side. Choosing the latter possibility, we find that
the experimental result corresponding to our eq.~(\ref{main.num}) is
\be
{\rm BR}[\bar{B} \to X_s \gamma]^{\rm exp}_{E_{\gamma} > 1.6~{\rm GeV}}
= \f{3.60}{3.73} (3.23 \pm 0.42) \times 10^{-4}
= (3.12 \pm 0.41) \times 10^{-4},
\ee
where the ratio of eqs.~(\ref{main.num}) and (\ref{br.tot}) has been
used as the rescaling factor. The difference between theory and
experiment can  now be written as
\mathindent0cm
\be
{\rm BR}_{\gamma}^{\rm th}- {\rm BR}_{\gamma}^{\rm exp} = (3.60 \pm 0.30) \times 10^{-4}
- (3.12 \pm 0.41) \times 10^{-4} = (0.48 \pm 0.50) \times 10^{-4},
\ee
\mathindent1cm
which is compatible with zero. Based on the present experimental
information, we have shown that the Two--Higgs--Doublet--Model II with
a charged Higgs boson lighter than 350~GeV is strongly disfavoured.

\section*{Acknowledgements}

We would like to thank U.~Aglietti, M.~Beneke, G.~Buchalla,
P.~Chankowski, U.~Haisch, A.~Hoang, G.~Isidori, Z.~Ligeti, M.~Luke,
G.~Martinelli and N.~Uraltsev for helpful discussions.  M.M.  has been
supported in part by the Polish Committee for Scientific Research
under grant 2~P03B~121~20, 2001-2003.

\newappendix{Appendix A}
\def\theequation{A.\arabic{equation}}
 
In this appendix, our numerical input parameters are collected. From
ref.~\cite{PData00}, we take \linebreak
$M_Z = 91.1882$~GeV, 
$M_W = 80.419$~GeV,
$\alpha_{em} = 1/137.036$, 
$\al(M_Z) = 0.1185 \pm 0.0020$,
$m_c^{\overline{\rm MS}}(m_c^{\overline{\rm MS}}) = (1.25 \pm 0.10)$~GeV
and
BR$[\bar{B} \to X_c e \bar \nu]_{\rm exp} = 0.1045 \pm 0.0021$.  

For the $b$-quark mass in the so-called 1S-scheme, we use $m_b^{1S} =
(4.69 \pm 0.03)$~GeV \cite{H00}. As far as the top-quark mass is
concerned,
$m_t^{\overline{\rm MS}}(m_t^{\overline{\rm MS}}) = (165 \pm 5)$~GeV is used,
which corresponds to $m_t^{\rm pole} = 174.3 \pm 5.1$~GeV \cite{PData00}.

In several places, definitions of  $m_t$ and $m_b$ have been left
unspecified, because it would become relevant only at higher orders of
perturbation theory. In such places, 
$m_t^{\overline{\rm MS}}(m_t^{\overline{\rm MS}})$ and $m_b^{1S}$ are
used.

When the $\mu$-dependence is being tested, each scale is made to vary from half
to twice its central value. The central value of $\mu_b$ is chosen to
be $m_b$. For the matching scale $\mu_0$, we take the central value of
$M_W$ in $K_c$ and in the quantities for which no flavour splitting is
performed. In $K_t$ and in the mass ratio $r(\mu_0)$, the central
value of $\mu_0$ is set to $m_t$.

The final result (\ref{main.num}) for the branching ratio (\ref{main})
has been found by using the above inputs as well as  the
quantities specified in eqs.~
(\ref{CKMratio}), 
(\ref{ewcor.num}), 
(\ref{ndel.num}), 
(\ref{mratio.num}),
(\ref{phase.num}) 
and
(\ref{ncb.num}).

\newappendix{Appendix B}
\def\theequation{B.\arabic{equation}}

Here, we determine the mass ratio 
$r(\mu_0) = m_b(\mu_0)^{\overline{\rm MS}}/m_b^{1S}$.
The NLO result reads
\mathindent0cm
\bea \label{mratio.NLO}
r_{\scs NLO}(\mu_0) = \left(\f{\al(\mu_0)}{\al(m_b)}\right)^{\f{12}{23}} 
\left\{ 1 + \f{\al(m_b)}{4\pi} \left[ \f{7462}{1587} \; \f{\al(\mu_0)}{\al(m_b)}
- \f{15926}{1587} \right] + \f{2}{9} \al(m_b)^2 \right\}
~\bbuildrel{=}_{\mu_0=m_t}^{}~ 0.611. 
\hspace{-7mm} \nonumber\\[-3mm] \\[-12mm] \nonumber
\eea
\mathindent1cm
The ${\cal O}(\al^2)$ term is included along the lines of the $\Upsilon$
  expansion \cite{HLM99}. Its effect on $r$ is at the level of 1\%
  only, and it is offset by an analogous term in $K_t$ (\ref{Kt}).

Since $m_b(\mu_0)$ could, in principle, be determined from a
high-energy observable (e.g. the Higgs width), we are allowed to
calculate the ratio $r(\mu_0)$ using as many orders of perturbation
theory as are currently known, i.e. up to ${\cal O}(\epsilon^3)$ in
the $\Upsilon$ expansion.

From eq. (168) of ref.~\cite{H00}, we find
\be \label{mratio.bs}
\f{ m_b^{\overline{\rm MS}}(m_b^{\overline{\rm MS}})}{m_b^{1S}} \approx 0.889
\left( \f{ \al(M_Z)}{0.1185} \right)^{-0.26}
\left( \f{ m_b^{1S}}{4.69} \right)^{0.04}
\left( \f{ m_c(m_c)}{1.25} \right)^{-0.003}
\left( \f{ \mu_b }{4.69} \right)_{\cdot}^{0.006}
\ee
All the masses and renormalization scales in
eqs.~(\ref{mratio.bs})--(\ref{mratio.num}) are expressed in GeV. The
dependence on $\mu_b$ in eq.~(\ref{mratio.bs}) originates only from
the fact that $\al(\mu_b)$ rather than $\al(m_b)$ has been used in
evaluating this relation. Varying $\mu_b$ by a factor of 2 around
4.69~GeV gives us an estimate of $\pm0.4\%$ for the neglected
higher-order effects.

The four--loop RGE for the quark mass imply
\be \label{mratio.tb}
\f{ m_b^{\overline{\rm MS}}(m_t)}
  { m_b^{\overline{\rm MS}}(m_b^{\overline{\rm MS}})} \approx 0.650
\left( \f{ \al(M_Z)}{0.1185} \right)^{-0.72}
\left( \f{ m_b^{\overline{\rm MS}}(m_b^{\overline{\rm MS}})}{4.17} \right)^{0.19}
\left( \f{ m_t}{165} \right)_{,}^{-0.08}
\ee
where the higher-order uncertainty is negligible (below 0.2\%).
Combining eqs.~(\ref{mratio.bs}) and (\ref{mratio.tb}), we find
\mathindent0cm
\be 
r(m_t) \equiv \f{ m_b^{\overline{\rm MS}}(m_t)}
  { m_b^{1S}} \approx 0.578 
\left( \f{ \al(M_Z)}{0.1185} \right)^{-1.0}
\left( \f{ m_b^{1S}}{4.69} \right)^{0.23}
\left( \f{ m_c(m_c)}{1.25} \right)^{-0.003}
\left( \f{ m_t}{165} \right)^{-0.08}
\left( \f{ \mu_b }{4.69} \right)_{\cdot}^{0.006}~~
\label{mratio.num}
\ee
\mathindent1cm
The above approximate formula works with better than 0.2\% accuracy
(compared with the complete one\footnote{
We thank A.~Hoang for sending us the numerical data from which
eq. (168) of ref.~\cite{H00} was derived.}
) when the input parameters vary within their errors (see
appendix A).

It is important to note that the central value in
eq.~(\ref{mratio.num}) is 5.4\% less than $r_{\scs NLO}$ in
eq.~(\ref{mratio.NLO}). The main reason of this suppression is the
$-3.2\%$ higher-than-NLO correction to
$m_b^{\overline{\rm MS}}( m_b^{\overline{\rm MS}})/m_b^{1S}$ 
(see eq.~(99) of ref.~\cite{H00}).  Switching from two-loop to four-loop
RGE causes a reduction of $r$ by only $0.7\%$. Additional effects
originate from the fact that the scales $\mu = m_b^{1S}$ and $\mu =
m_b(m_b)$ have been identified in eq.~(\ref{mratio.NLO}).

The ratio $r(\mu_0)$ depends only logarithmically on the actual value
of $m_b^{1S}$. This is reflected in eq.~(\ref{mratio.num}) where the
power of $m_b^{1S}/4.69$ is equal to 0.23 only. Consequently, our
results are not very sensitive to the numerical input we use for
$m_b^{1S}$. They would not change much if instead of $m_b^{1S}$ we
used another definition of the "kinetic" mass of the b-quark, e.g. the
ones defined in refs.~\cite{BSU97,BS99}, or the $\overline{\rm MS}$
mass at appropriately chosen $\mu$. However, in such a case,
eq.~(\ref{Kt}) would need to be modified accordingly.

\newappendix{Appendix C}
\def\theequation{C.\arabic{equation}}

The present appendix is devoted to a determination of the semileptonic
phase-space factor:
\be \label{phase2}
C = \left| \f{V_{ub}}{V_{cb}} \right|^2 
\f{\Gamma[\bar{B} \to X_c e \bar{\nu}]}{\Gamma[\bar{B} \to X_u e \bar{\nu}]}
\ee
along the lines of the $\Upsilon$ expansion \cite{HLM99}.

We begin with the NNLO expressions for both decay rates expressed in
terms of pole quark masses and $\al \equiv \al^{(n_f=5)}(m_b) \approx 0.219$:
\bea 
\label{crate}
\Gamma[\bar{B} \to X_c e \bar{\nu}] &=& \f{G_F^2 (m_b^{\rm pole})^5}{192\pi^3} |V_{cb}|^2
g(z_p) \left[ 1 + \epsilon \f{\al}{\pi} p_c^{(1)}(z_p) 
              + \epsilon^2 \f{\al^2}{\pi^2} p_c^{(2)}(z_p) 
\right. \nonumber\\ && \hspace{44mm} \left.
              + \f{\lambda_1}{2 m_b^2} 
              + \f{\lambda_2}{m_b^2} \left( \f{3}{2} - \f{6 (1-z_p)^4}{g(z_p)} \right) \right],\\
\Gamma[\bar{B} \to X_u e \bar{\nu}] &=& \f{G_F^2 (m_b^{\rm pole})^5}{192\pi^3} |V_{ub}|^2
\left[ 1 + \epsilon \f{\al}{\pi} p_u^{(1)} 
              + \epsilon^2 \f{\al^2}{\pi^2} p_u^{(2)}(z_p) 
              + \f{\lambda_1}{2 m_b^2} - \f{9\lambda_2}{2m_b^2} \right],
\label{urate} 
\eea
where $z_p = (m_c^{\rm pole}/m_b^{\rm pole})^2$ and $g(z) = 1 - 8 z + 8
z^3 - z^4 - 12 z^2 \ln z$. The auxiliary parameter $\epsilon \equiv 1$
will be helpful below in denoting orders of the $\Upsilon$ expansion.

The perturbative corrections are as follows \cite{ref_C_pert}:
\bea
p_c^{(1)}(z) &=& -\f{2 h(z)}{3 g(z)}, \label{pc1}\\
p_c^{(2)}(z=0.09) &=& -1.68 \beta_0^{(4)} + (1.4 \pm 0.4) = -12.6 \pm 0.4,\label{pc2}\\
p_u^{(1)} &=& \f{25}{6} - \f{2}{3} \pi^2, \label{pu1}\\
p_u^{(2)}(z) &=& -3.22 \beta_0^{(4)} + 5.53 + f_u(z)
= -21.30 + f_u(z), \label{pu2}
\eea
where $\beta_0^{(n_f)} = 11 - \f{2}{3} n_f$~ and
\mathindent0cm
\bea
h(z) &=&
- (1-z^2) \left( \f{25}{4}- \f{239}{3} z + \f{25}{4} z^2 \right) 
+ z \ln z \left( 20 + 90 z - \f{4}{3} z^2 + \f{17}{3} z^3 \right) 
+ z^2 \ln^2 z \; ( 36 + z^2) 
\nonumber \\ && 
+ (1-z^2) \left( \f{17}{3} - \f{64}{3} z + \f{17}{3} z^2 \right) \ln (1-z) 
- 4 ( 1 + 30 z^2 + z^4 ) \ln z \; \ln (1-z) 
\nonumber \\ && \hspace{-17mm}
- (1 + 16z^2 + z^4) [ 6 {\rm Li}_2(z) - \pi^2 ] 
- 32 z^{3/2} (1+z) \left[ \pi^2 
     - 4 {\rm Li}_2(\sqrt{z}) +  4 {\rm Li}_2(-\sqrt{z}) 
     - 2 \ln \left( \f{1-\sqrt{z}}{1+\sqrt{z}} \right) \ln z \right]. 
\nonumber \eea 
As far as the charm-mass correction $f_u(z)$ in eq.~(\ref{pu2}) is
concerned, the only thing we know for sure is that $f_u(0)=0$. A rough
estimate of $f_u(0.09)$ can be obtained by considering the relation
between pole and $\overline{\rm MS}$ bottom masses, for which
charm-loop corrections are known at two \cite{GBGS90} and three
\cite{H00} loops. Similarly to $f_u(z)$, they originate from charm
loop insertions on gluon lines.  For $m_c/m_b=0.3$, we learn from
ref.~\cite{H00} that two-loop charm mass effects lead to an extra term
$ 0.43 (\alpha_s/\pi)^2$. As the charm loops are closely related to
the BLM corrections, which amount to $1.56 \beta_0 (\alpha_s/\pi)^2$
in the relation between masses, we can rescale them by a factor
$-3.22/1.56$ to obtain $f_u(0.09)\sim -0.9$.  Of course, this is only
a very rough estimate, and an actual calculation of $f_u(z)$ would be
welcome.

When the ratio (\ref{phase2}) is calculated, the overall factors
of $(m_b^{\rm pole})^5$ cancel. The remaining explicit dependence on
pole quark masses can be eliminated with the help of the following
relation:
\be \label{rcb}
\f{m_c^{\rm pole}}{m_b^{\rm pole}} 
= 1-\f{m_b^{\rm pole}-m_c^{\rm pole}}{m_b^{\rm pole}}
= 1-\f{\overline{m}_B-\overline{m}_D 
        + \f{\lambda_1}{2\overline{m}_B}- \f{\lambda_1}{2\overline{m}_D}
        + {\cal O}\left(\f{\Lambda_{\rm QCD}^3}{\overline{m}_D^2}\right)
}{
      \f{1}{2} m_{\Upsilon} \left( 1-\f{\Delta^2}{(m_{\Upsilon}/2)^2} \right)
\left\{ 1 + \epsilon \f{(\al C_F)^2}{8} \left[ 1 + \epsilon \f{\al}{\pi} Y 
+ {\cal O}(\al^2) \right] \right\}},
\hspace{4mm}
\ee
\mathindent1cm
with $C_F=\f{4}{3}$, $m_{\Upsilon} \approx 9.460$~GeV,
$\overline{m}_B = \f{1}{8} ( 3 m_{B^{\star\pm}} + m_{B^{     \pm}} 
                           + 3 m_{B^{\star 0}}  + m_{B^0}) \approx 5.314$~GeV
and
$\overline{m}_D = \f{1}{8} ( 3 m_{D^{\star\pm}} + m_{D^{     \pm}} 
                           + 3 m_{D^{\star 0}}  + m_{D^0}) \approx 1.973$~GeV.
In the denominator, $\Delta$ describes the non-perturbative
contribution responsible for the difference between $\f{1}{2}
m_{\Upsilon}$ and $m_b^{1S}$. From ref.~\cite{H00}, one
finds\footnote{
Here, we neglect the ${\cal O}(\al^3)$ difference between
$\al^{(n_f=4)}(m_b)$ and $\al^{(n_f=5)}(m_b)$ in the $\overline{\rm MS}$ scheme.}
\bea
\f{\Delta^2}{(m_{\Upsilon}/2)} &=& (0.04 \pm 0.03)~{\rm GeV}, \label{Dt}\\
Y &=& \f{203}{18} - \f{25}{3} \ln (\al C_F) + y\left( \f{3 m_c}{\al m_{\Upsilon}}\right)
\label{yy}
\eea
with
\be
y(a) = \f{2}{3} \ln\f{a}{2} +\f{\pi}{2}a -\f{4}{3}a^2 +\f{2\pi}{3}a^3 
        + \f{4-2a^2-8a^4}{3\sqrt{a^2-1}} \arctan \f{\sqrt{a-1}}{\sqrt{a+1}}.
\ee
Below, $m_c(m_c) \approx 1.25$~GeV \cite{PData00} will be used in this
function. The influence of $y(a)$ on $C$ is less than 0.3\%, so it does
not matter what definition of $m_c$ is chosen here.

In the following, we shall make use of eq.~(\ref{rcb}), and expand all
the functions of $z_p$ in eqs.~(\ref{crate}) and (\ref{urate}) around
\be \label{z0}
z_0 = \left( 1-\f{\overline{m}_B-\overline{m}_D}{m_{\Upsilon}/2} \right)^2 \approx 0.0863,
\ee
neglecting ${\cal O}(\epsilon^3)$, ${\cal O}(\al^4)$, ${\cal
O}(\Lambda_{\scs QCD}^3/m_Q^3)$ and ${\cal O}(\epsilon \times
\Lambda_{\scs QCD}^2/m_Q^2)$. As argued in ref.~\cite{HLM99}, the
leading renormalons cancel when such an expansion is performed for
physical observables. Indeed, the QCD perturbation series converges
remarkably well in our final result for the ratio (\ref{phase2}):
\mathindent0cm
\bea
C &=& g(z_0) \left\{ 1 + \epsilon \left[ \f{\al}{\pi} 
\left( p_c^{(1)}(z_0)-p_u^{(1)} \right) + \kappa \f{(\al C_F)^2}{8} \right]
\right. \nonumber \\[1mm] &+& \left. \epsilon^2 \left[ \f{\al^2}{\pi^2} 
\left( p_c^{(2)}(z_0)-p_u^{(2)}(z_0) + p_u^{(1)}(z_0)^2 - p_c^{(1)}(z_0) p_u^{(1)} \right) 
+\kappa \f{(\al C_F)^2 \al}{8 \pi} \left( Y -p_u^{(1)} -\f{2h'(z_0)}{3g'(z_0)}\right)\right]
\right. \nonumber \\[1mm] && \left. \hspace{15mm}
+ \f{\kappa \lambda_1}{2 \overline{m}_B \overline{m}_D}
+ \f{24 \lambda_2}{m_{\Upsilon}^2} \left( 1 - \f{(1-z_0)^4}{g(z_0)}\right)
- \f{\kappa \Delta^2}{(m_{\Upsilon}/2)^2} \right\} \nonumber\\
&\approx& 0.534 \left[ 1 + 0.019 \epsilon + (0.001 - 0.005 f_u(z_0) \pm 0.002 ) \epsilon^2
-\f{\lambda_1}{7.1~{\rm GeV}^2} -\f{\lambda_2}{12~{\rm GeV}^2} 
+\f{\Delta^2}{7.6~{\rm GeV}^2} \right] \nonumber\\[2mm]
&=& 0.575 \times ( 1 \pm 0.01_{\rm pert} \pm 0.02_{\lambda_1} \pm 0.02_{\Delta} ), 
\label{phase.num}
\eea
\mathindent1cm
where
\be
\kappa = 2 \sqrt{z_0}\left(1-\sqrt{z_0}\,\right) \f{g'(z_0)}{g(z_0)} \approx 2.94.
\ee
In the middle step, the error in the ${\cal O}(\epsilon^2)$ term
originates only from $p_u^{(2)}(z_0) = -12.6 \pm 0.4$ (see
eq.~(\ref{pc2})). In the last step, $f_u(z_0)$ is set to $-0.9$, and
an overall $\pm 1\%$ perturbative uncertainty is assumed, resulting
from the fact that our estimate of $f_u(z_0)$ is very rough, from the
error in $p_u^{(2)}(z_0)$ and from missing perturbative higher
orders. As far as the non-perturbative parameters are concerned, we
use $\lambda_1 = (-0.27 \pm 0.10 \pm 0.04)~{\rm GeV}^2$~\cite{HLM99},\footnote{
It is consistent with $\lambda_1 = (-0.196 \pm 0.065 \pm 0.072)~{\rm GeV}^2$
recently extracted by CLEO \cite{CLEO} from the $\bar{B} \to X_s \gamma$ spectrum.}
$\lambda_2 = 0.12~{\rm GeV}^2$ and eq.~(\ref{Dt}).

A delicate point in our calculation of $C$ is the fact
that the unknown ${\cal O}(\Lambda_{\scs QCD}^3/\overline{m}_D^2)$
corrections have been neglected in the numerator of
eq.~(\ref{rcb}). Their potential effect can be studied by replacing
$\lambda_1 \to \lambda_1 \pm {\cal O}(\Lambda_{\scs
QCD}^3)/\overline{m}_D$ in eq.~(\ref{phase.num}). However, if such
corrections were sizeable, they would affect the determination of
$\lambda_1$ from the semileptonic spectrum in ref.~\cite{HLM99}. In the
present paper, we assume that those corrections are included in the
error of $\lambda_1$ (and in its central value). A further study of
this point is warranted.

        It is worth mentioning that our final central value $C=0.575$
can be reproduced from the ratio of eqs.~(\ref{crate}) and
(\ref{urate}) expanded in $\al$ and $\lambda_i$, for $z_p=0.285^2$ at
NLO, and $z_p=0.298^2$ at NNLO. Those two values are consistent with
$z_p= (0.29 \pm 0.02)^2$ which has been used in many previous analyses
of $\bar{B} \to X_s \gamma$. On the other hand, when $z_p = (0.29 \pm
0.02)^2$ is used in such a ratio, one obtains at NNLO
\be
C_{\rm direct} = 0.595 \times (1 \pm 0.08 \pm 0.05),
\ee
where the first error comes from $z_p$, and the second one from poor
convergence of the perturbation series. The uncertainties here are
much larger than in the result (\ref{phase.num}) obtained with the help of
the $\Upsilon$ expansion.

\newappendix{Appendix D}
\def\theequation{D.\arabic{equation}}

\begin{figure}[htb]
\begin{center}
\includegraphics[width=14cm,angle=0]{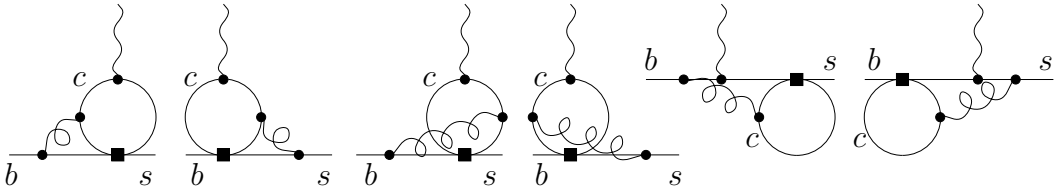}
\caption{\sf Leading contributions to the $b \to s \gamma$ matrix elements of
$P_1$ and $P_2$.}
\label{fig:GHWdiags}
\end{center}
\ \\[-36mm]
\hspace*{10cm}$b$\hspace{22mm}$s$\hspace{4mm}$b$\hspace{21mm}$s$\\[-3mm]
\hspace*{24mm}$c$\hspace{13mm}$c$\hspace{3cm}$c$\hspace{13mm}$c$\\[3mm]
\hspace*{113.5mm}$c$\hspace{12.5mm}$c$\\
\hspace*{15mm}$b$\hspace{16mm}$s$\hspace{5mm}$b$\hspace{15mm}$s$
  \hspace{3mm}$b$\hspace{14mm}$s$\hspace{5mm}$b$\hspace{16mm}$s$\\[-1mm]
\end{figure}

In this appendix, we present the analytical formulae and calculate the
numerical values of the functions $a(z)$ and $b(z)$ that occur in the
expression for $K_c$ (\ref{Kc}). Those two functions of $z =
(m_c/m_b)^2$ originate from the two--loop $b \to s \gamma$ matrix
elements of the 4-quark operators $P_1$ and $P_2$ (\ref{ops}). The
relevant diagrams are shown in fig.~\ref{fig:GHWdiags}. They were
calculated in ref.~\cite{GHW96}. Their colour structure implies that
$\me{P_1} = -\f{1}{6} \me{P_2}$. Additive constants in the functions
$a(z)$ and $b(z)$ are chosen in such a way that both functions vanish
at $z=0$. Explicitly,
\mathindent0cm
\bea \ \nonumber\\[1cm] \ \\[-22mm] 
\begin{array}{rcl}
a(z)  &=& \f{16}{9} \left\{ 
\left[ \f{5}{2} -\f{1}{3}\pi^2 -3 \zeta(3) + \left( \f{5}{2} - \f{3}{4} \pi^2 \right) L
+ \f{1}{4} L^2 + \f{1}{12} L^3 \right] z 
\right. \\[2mm] && \hspace{3mm} \left.
+\left[\f{7}{4} +\f{2}{3} \pi^2 -\f{1}{2} \pi^2 L -\f{1}{4} L^2 +\f{1}{12} L^3 \right] z^2 + 
\left[ -\f{7}{6} -\f{1}{4} \pi^2 + 2L - \f{3}{4}L^2 \right] z^3
\right. \\[2mm] && \hspace{3mm} \left.
+ i \pi \left[ \left( 2 -\f{1}{6} \pi^2 + \f{1}{2} L + \f{1}{2} L^2 \right) z 
+ \left( \f{1}{2} -\f{1}{6} \pi^2 - L + \f{1}{2} L^2 \right) z^2 + z^3 \right] \right\}
+ {\cal O}(z^4 L^4),
\\[3mm] 
b(z) &=& \hspace{-2mm} -\frac{8}{9} \left\{
\left( -3 +\f{1}{6} \pi^2 - L \right) z 
+ \left( \f{1}{2} + \pi^2 - 2 L - \f{1}{2} L^2 \right) z^2 
+ \left(-\f{25}{12} -\f{1}{9} \pi^2 - \f{19}{18} L + 2 L^2 \right) z^3 
\right. \\[2mm] && \hspace{3mm} \left. 
- \f{2}{3}\pi^2 z^{3/2}
+ i \pi \left[ -z + (1 - 2 L) z^2 + \left(-\f{10}{9} + \f{4}{3} L \right) z^3 \right] \right\}
+ {\cal O}(z^4 L^4),
\end{array} \nonumber\\[-6.5mm] \eea
\mathindent1cm
where $L = \ln z$. 

The imaginary parts of $a(z)$ and $b(z)$ have very little influence on
our prediction for BR$_\gamma$. Since the LO amplitude is real, they
affect the r.h.s.~of eq.~(\ref{Pdel}) only via ${\cal O}(\al^2)$ terms
that we neglect anyway\footnote{
Except for the ${\cal O}(\al^2)$ contributions to the ratio $r(\mu_0)$.}
and via the very small ${\cal O}(V_{ub})$ correction in eq.~(\ref{Kc}).
If we included the imaginary parts of $K_c$ and $K_t$, our final result
(\ref{main.num}) would get enhanced by only around 0.5\%.\\[-5mm]

As we have mentioned in the introduction, the choice of
renormalization scheme for $m_c$ and $m_b$ in $\me{P_{1,2}}$ is very
important for BR$_\gamma$. In principle, this choice is a NNLO issue
that can be resolved only after calculating three-loop corrections to
the diagrams of fig.~\ref{fig:GHWdiags}. Since calculating finite
parts of such diagrams would be a very difficult task at present, we
have to guess what the optimal choice of $m_c$ and $m_b$ is, on the
basis of our experience from other calculations.

        All the factors of $m_c$ in $\me{P_{1,2}}$ originate from
explicit mass factors in the charm-quark propagators. In the real
parts of $\me{P_{1,2}}$, those charm quarks are dominantly off-shell,
with momentum scale $\mu$ set by $m_b$. Actually, we are not able to
decide whether this scale is $m_b$, $\f{1}{2}m_b$ or
$\f{1}{3}m_b$. Therefore, we shall vary $\mu$ between $m_c \sim
\f{1}{3}m_b$ and $m_b$, and use $m_c^{\overline{\rm MS}}(\mu)$ in
$a(z)$ and $b(z)$.

        As far as the factors of $m_b$ are concerned, they originate
either from the overall momentum release in $b \to s \gamma$ or from the
explicit appearance of $m_b$ in the $b$-quark propagators. In the
first case, the appropriate choice of $m_b$ is a low-virtuality mass.
In the second case, there is no intuitive argument that  could tell us
whether $m_b^{\rm pole}$ or $m_b(m_b)$ is preferred. However, we think
that as long as the three-loop diagrams remain unknown, the best
choice is to set all the factors of $m_b$ equal to
$m_b^{1S}$ in $a(z)$ and $b(z)$. The distinction between $m_b^{1S}$
and $m_b^{\rm pole}$ as well as the infrared sensitivity of the pole
mass can be ignored here, because the uncertainties due to $m_c(\mu)$
are very large.

        In determining the optimal renormalization scheme for $m_b$ in
the ratio $m_c^2/m_b^2$, we have to take into account that the
considered perturbative amplitude arises as a particular term in the
operator product expansion for the inclusive decay of the $\bar{B}$
meson.  Thus, the ratio $m_c^2/m_b^2$ should be understood as
$m_c^2/Q^2$, where $Q^2$ is the squared "mean" momentum of the
$b$-quark inside the $\bar{B}$-meson. Its value is given by one of the
"kinetic" b-quark masses \cite{HLM99,BSU97,BS99},
e.g. $(m_b^{1S})^2$. Given the large error in $m_c(\mu)$, it does not
matter which definition of the bottom kinetic mass is chosen here.

        It remains to determine the numerical value of 
$m_c^{\overline{\rm MS}}(\mu)/m_b^{1S}$. 
From $m_c = m_c(m_c) = (1.25 \pm 0.10)$~GeV \cite{PData00}
and the RGE for $m_c(\mu)$, we find 
\be \label{ncb.num}
m_c^{\overline{\rm MS}}(\mu)/m_b^{\rm 1S} = 0.22 \pm 0.04,
\ee
which implies
\bea
a(z) &=& (0.97 \pm 0.25) \;+\; i ( 1.01 \pm 0.15 ),\\
b(z) &=& (-0.04 \pm 0.01) \;+\; i ( 0.09 \pm 0.02 ),
\eea
for $\mu$ varying between $m_c$ and $m_b$. The uncertainty in
$m_c(m_c)$ is practically irrelevant here, when compared to the
dominant error that originates from a variation of $\mu$.  We also
note that our central value $m_c/m_b = 0.22$, is very close to the
RG-invariant $\overline{\rm MS}$ ratio $m_c(\mu)/m_b(\mu)=0.215$.

\newappendix{Appendix E}
\def\theequation{E.\arabic{equation}}

Here, we discuss the function $B(E_0)$ in eq.~(\ref{Pdel}) that arises
from  $b \to s \gamma g$ and  $b \to s \gamma q \bar{q}$ with
$q=u,d,s$:
\be \label{bdel}
B(E_0) = \f{\al(\mu_b)}{\pi} 
\sum_{\begin{array}{c} \ \\[-7mm] {\scriptscriptstyle i,j = 1, ..., 8} \\[-2mm] 
                         {\scriptscriptstyle i \leq j} \end{array}} 
C_i^{(0)\rm eff}(\mu_b) \;\; C_j^{(0)\rm eff}(\mu_b) \;\; \phi_{ij}
\left( 1 - \f{2 E_0}{m_b} \right)
\;+\; \beta_{q\bar{q}}(E_0).
\ee
The functions $\phi_{ij}(\delta)$ originate from the gluon
bremsstrahlung $b \to s \gamma g$ \cite{AG95,P96,LLMW99}. For $i,j \in
\{1,2,7,8\}$ their explicit form is as follows:
\mathindent0cm
\bea
\phi_{22}(\delta) &=& \f{16 z}{27} \left[ 
\delta \int_0^{(1-\delta)/z} dt \; (1-zt) \left| \f{G(t)}{t} + \f{1}{2} \right|^2 
\;+\;
\int_{(1-\delta)/z}^{1/z} dt \; (1-zt)^2  \left| \f{G(t)}{t} + \f{1}{2} \right|^2 
\right], \label{phi22}\\
\phi_{27}(\delta) &=& -\f{8 z^2}{9} \left[ 
\delta \int_0^{(1-\delta)/z} dt\;    {\rm Re} \left( G(t) + \f{t}{2} \right) \;+\;
\int_{(1-\delta)/z}^{1/z} dt\;(1-zt) {\rm Re} \left( G(t) + \f{t}{2} \right) \right],
\label{phi27}\\
\phi_{77}(\delta) &=& -\f{2}{3} \ln^2 \delta -\f{7}{3} \ln \delta - \f{31}{9} + 
\f{10}{3} \delta + \f{1}{3} \delta^2 - \f{2}{9} \delta^3
+ \f{1}{3} \delta ( \delta - 4 ) \ln \delta, \label{phi77}\\
\phi_{78}(\delta) &=& \f{8}{9} \left[ {\rm Li}_2(1-\delta) - \f{\pi^2}{6} - \delta  
\ln \delta + \f{9}{4} \delta - \f{1}{4} \delta^2 + \f{1}{12} \delta^3 \right],\\
\phi_{88}(\delta) &=& \f{1}{27} \left\{ - 2 \ln \f{m_b}{m_s} 
                     \left[ \delta^2 + 2 \delta + 4 \ln(1-\delta) \right] 
\right. \nonumber \\ && \hspace{1cm} \left.
+4{\rm Li}_2(1-\delta) -\f{2\pi^2}{3} -\delta(2+\delta)\ln\delta + 
8\ln(1-\delta) -\f{2}{3} \delta^3 + 3 \delta^2 + 7 \delta \right\}, \label{phi88}
\eea
\be
\phi_{11}=\f{1}{36}\phi_{22},~~~~~~
\phi_{12}=-\f{1}{3}\phi_{22},~~~~~~
\phi_{17}=-\f{1}{6}\phi_{27},~~~~~~
\phi_{18}=\f{1}{18}\phi_{27},~~~~~~
\phi_{28}=-\f{1}{3}\phi_{27}.
\ee
\mathindent1cm
They are identical to the $f_{ij}(\delta)$ used in many previous analyses
of $\bar{B} \to X_s \gamma$, except for the case $i=j=7$.  The
difference between $\phi_{77}(\delta)$ and $f_{77}(\delta)$ from
eq.~(37) of ref.~\cite{CMM97} is given by the first three terms in
eq.~(\ref{phi77}): the Sudakov logarithms ($\ln^2\delta$, $\ln\delta$)
and a constant term that makes $\phi_{77}(\delta)$ vanish at $\delta
\to 1$.  For simplicity, we refrain from resumming the Sudakov
logarithms here, because we are interested only in energy cut--offs $E_0
< 2.1$~GeV ($\delta > 0.1$), for which $| \phi_{77}(\delta) | < 1$,
i.e. the logarithmic divergence of $\phi_{77}(\delta)$ at $\delta \to
0$ is not yet relevant.

The function $G(t)$ that appears in eqs.~(\ref{phi22}) and
(\ref{phi27}) reads
\be
G(t) = \left\{ \begin{array}{cc} 
- 2 \arctan^2 \sqrt{ t/(4-t)}, & \mbox{ for $t < 4$} \vspace{0.2cm} \\
-\pi^2/2 + 2 \ln^2[(\sqrt{t} + \sqrt{t-4})/2] 
- 2 i \pi \ln[(\sqrt{t} + \sqrt{t-4})/2], & \mbox{ for $t \geq 4$}. 
\end{array} \right.
\ee
In our numerical analysis, the parameter $z=m_c^2/m_b^2$ entering those
equations is set equal to $(0.22 \pm 0.04)^2$, as determined in appendix
D. In eq.~(\ref{phi88}), we follow ref.~\cite{CMM97} and use
$m_b/m_s=50$.

The functions $\phi_{ij}$ with $3 \leq (i~{\rm or}~j) \leq 6$
have only a 0.1\% effect on 
BR$[\bar{B} \to X_s \gamma]_{E_{\gamma}>1.6\;{\rm GeV}}$. 
Therefore, we shall not give them explicitly here. They can be read
from the results of ref.~\cite{P96}.

The coefficients $C_i^{(0)\rm eff}(\mu_b)$ are equal to those in eqs.~(20)
and (22) of ref.~\cite{CMM97}. They read ($C_i = C_i^{\rm eff}$ for $i=1,...,6$)
\mathindent0cm
\bea
\begin{array}{rcrll}
C^{(0)}_1(\mu_b) &=&           \eta^{a_3} &            -\eta^{a_4},&\\[2mm]
C^{(0)}_2(\mu_b) &=& \f{2}{3}  \eta^{a_3} &   +\f{1}{3} \eta^{a_4},&\\[2mm]
C^{(0)}_3(\mu_b) &=& \f{2}{63} \eta^{a_3} & -\f{1}{27}  \eta^{a_4} &\hspace{-3mm}
                - 0.0659 \,\eta^{a_5}  + 0.0595 \,\eta^{a_6} 
                - 0.0218 \,\eta^{a_7}  + 0.0335 \,\eta^{a_6},\\[2mm]
C^{(0)}_4(\mu_b) &=& \f{1}{21} \eta^{a_3} &   +\f{1}{9} \eta^{a_4} &\hspace{-3mm}
                + 0.0237 \,\eta^{a_5}  - 0.0173 \,\eta^{a_6} 
                - 0.1336 \,\eta^{a_7}  - 0.0316 \,\eta^{a_8},\\[2mm]
C^{(0)}_5(\mu_b) &=&-\f{1}{126}\eta^{a_3} & +\f{1}{108} \eta^{a_4} &\hspace{-3mm} 
                + 0.0094 \,\eta^{a_5}  - 0.0100 \,\eta^{a_6} 
                + 0.0010 \,\eta^{a_7}  - 0.0017 \,\eta^{a_8},\\[2mm]
C^{(0)}_6(\mu_b) &=&-\f{1}{84} \eta^{a_3} &  -\f{1}{36} \eta^{a_4} &\hspace{-3mm}
                + 0.0108 \,\eta^{a_5}  + 0.0163 \,\eta^{a_6} 
                + 0.0103 \,\eta^{a_7}  + 0.0023 \,\eta^{a_8},\\[2mm]
C^{(0)\rm eff}_7(\mu_b) &=& 
\multicolumn{3}{l}{
\left( -\f{4}{3} F_0^t(x)  + \f{42678}{30253} \right) \eta^{a_1} +
\left( -\f{1}{2} A^t_0(x) +\f{4}{3} F_0^t(x)  - \f{86697}{103460} \right) \eta^{a_2}}\\[3mm]
&& -\f{3}{7} \eta^{a_3} & -\f{1}{14} \eta^{a_4} &\hspace{-3mm}
                - 0.6494 \,\eta^{a_5}  - 0.0380 \,\eta^{a_6} 
                - 0.0185 \,\eta^{a_7}  - 0.0057 \,\eta^{a_8},\\[2mm]
C^{(0)\rm eff}_8(\mu_b) &=& 
\multicolumn{2}{l}{\left( -\f{1}{2} F_0^t(x)  + \f{64017}{121012} \right) \eta^{a_1}}&
\hspace{-3mm}    -0.9135 \,\eta^{a_5}  +  0.0873 \,\eta^{a_6}       
                 -0.0571 \,\eta^{a_7}  +  0.0209 \,\eta^{a_8}, \hspace{5mm}
\end{array} \nonumber\\[-7mm]
\eea 
\mathindent1cm
where $x=m_t(m_t)^2/M_W^2$. The powers $a_i$ can be found in
table~\ref{tab:mag}, in section \ref{sec:NLO}.

It is important to mention that the sensitivity of $B(E_0)$ to $m_b$
via the argument of $\phi_{ij}$ is rather weak when $E_0 =
1.6$ GeV. When $m_b$ is varied between 4.5 and 4.9~GeV here, our final
result for BR$[\bar{B} \to X_s \gamma]_{E_{\gamma} > 1.6 \,{\rm GeV}}$ is
affected by only around 0.7\%. In section \ref{sec:num}, we have
neglected this uncertainty, and used $m_b = m_b^{1S}$ in
eq.~(\ref{bdel}).

The term denoted by $\beta_{q\bar{q}}(E_0)$ in eq.~(\ref{bdel}) stands
for the contributions from $b \to sq\bar{q}\gamma$ transitions with
$q=u,d,s$. Perturbatively, such effects on $\Gamma[b \to X_s \gamma]$
are suppressed by either 
\be
\left| \f{C^{(0)}_{3,...,6}(\mu_b)}{C^{(0)\rm eff}_7(\mu_b)} \right|^2 < 0.05 
\hspace{2cm} {\rm or} \hspace{2cm} 
\left| \f{V^*_{us} V_{ub}}{V_{ts}^* V_{tb}} 
       \f{C_2^{(0)}(\mu_b)}{C^{(0)\rm eff}_7(\mu_b)} \right|^2 \approx 0.002 
\ee
with respect to the leading terms. Further suppression occurs when we
restrict ourselves to high-energy photons \cite{LLMW99}. In our numerical analysis
we have set $\beta_{q\bar{q}}(E_0)$ to zero without including any
additional uncertainty, which we expect to be acceptable at present
for an energy cut--off $E_0 \geq 1.6$~GeV.

\setlength {\baselineskip}{0.2in}
 

\begin{thebibliography}{99}
\newcommand{\np}[3]{Nucl. Phys. {\bf B#1} (#2) #3}
\newcommand{\pl}[3]{Phys. Lett. {\bf B#1} (#2) #3}
\newcommand{\pr}[3]{Phys. Rev.  {\bf D#1} (#2) #3}
\newcommand{\prl}[3]{Phys. Rev. Lett. {\bf #1} (#2) #3}
\newcommand{\prp}[3]{Phys. Rept. {\bf #1} (#2) #3}
\newcommand{\zpc}[3]{Z. Phys. {\bf C#1} (#2) #3}

\bibitem{CDGG98} M.~Ciuchini, G.~Degrassi, P.~Gambino and G.F.~Giudice,
                \np{527}{1998}{21}, \\ {\it ibid.} {\bf B534} (1998) 3.
\bibitem{DGG00} G.~Degrassi, P.~Gambino and G.F.~Giudice, JHEP {\bf 0012} (2000) 009.
\bibitem{MPR98} M.~Misiak, S.~Pokorski and J.~Rosiek, hep-ph/9703442,
  published in the Review Volume ``Heavy Flavors II'', eds. A.J.~Buras
  and M.~Lindner, World Scientific Publishing Co., Singapore, 1998.
\bibitem{CLEO} S.~Chen {\it et al.} (CLEO Collaboration), hep-ex/0108032.
\bibitem{BELLE} H.~Tajima, talk given at the {\it 20th International
            Symposium on Lepton-Photon Interactions}, Rome, July 2001.
\bibitem{ALEPH} R.~Barate {\it et al.}, \pl{429}{1998}{169}.
\bibitem{1.ov.mc} 
  M.B.~Voloshin, \pl{397}{1997}{275};\\
  A.~Khodjamirian {\it et al.}, \pl{402}{1997}{167};\\
  Z.~Ligeti, L.~Randall and M.B.~Wise, \pl{402}{1997}{178};\\
  A.K.~Grant, A.G.~Morgan, S.~Nussinov and R.D.~Peccei, \pr{56}{1997}{3151};\\
  G.~Buchalla, G.~Isidori and S.J.~Rey, \np{511}{1998}{594}.
\bibitem{1.ov.mb} 
  A.F.~Falk, M.~Luke and M.J.~Savage, \pr{49}{1994}{3367}.
\bibitem{KLP95} A.~Kapustin, Z.~Ligeti and H.~D.~Politzer, \pl{357}{1995}{653}.
\bibitem{KN99} A.L.~Kagan and M.~Neubert, Eur.~Phys.~J. {\bf C7} (1999) 5.
\bibitem{GHW96}   C.~Greub, T.~Hurth and D.~Wyler, \pr{54}{1996}{3350};\\
  A.J.~Buras, A.~Czarnecki, M.~Misiak and J.~Urban, \np{611}{2001}{488}.
\bibitem{CMM97} K.~Chetyrkin, M.~Misiak and M.~M{\"u}nz, \pl{400}{1997}{206},\\
                                              {\it ibid.} {\bf B425} (1998) 414 (E).
\bibitem{AG95} A.~Ali and C.~Greub, \pl{361}{1995}{146}.
\bibitem{P96}  N.~Pott, \pr{54}{1996}{938}.
\bibitem{MM95} M.~Misiak and M.~M{\"u}nz, \pl{344}{1995}{308}.
\bibitem{match2} 
  K.~Adel and Y.P.~Yao, \pr{49}{1994}{4945};\\
  C.~Greub and T.~Hurth, \pr{56}{1997}{2934};\\
  A.J.~Buras, A.~Kwiatkowski and N.~Pott, \np{517}{1998}{353}.
\bibitem{CM98}   A.~Czarnecki and W.~Marciano, \prl{81}{1998}{277}.
\bibitem{BM00}  K.~Baranowski and M.~Misiak, \pl{483}{2000}{410}.
\bibitem{GH00}  P.~Gambino and U.~Haisch, JHEP {\bf  09} (2000) 001.
\bibitem{BKP97} A.J.~Buras, A.~Kwiatkowski and N.~Pott, \pl{414}{1997}{157},\\
                                               {\it ibid.} {\bf B434} (1998) 459 (E).
\bibitem{HLM99} A.H.~Hoang, Z.~Ligeti and A.V.~Manohar, \prl{82}{1999}{277},
                                                         \pr{59}{1999}{074017}.
\bibitem{BMMP94} A. J. Buras, M. Misiak, M. M{\"u}nz, and S. Pokorski,
                        \np{424}{1994}{374}.
\bibitem{BMU00.1} C.~Bobeth, M.~Misiak and J.~Urban, \np{574}{2000}{291}.
\bibitem{S81} A.~Sirlin, \np{196}{1982}{83}.
\bibitem{B98} C.~Bauer, \pr{57}{1998}{5611}.
\bibitem{CAFLMPRS00} M.~Ciuchini {\it et al.},
   JHEP {\bf 0107} (2001) 013. 
\bibitem{mele} S.~Mele,
talk at {\it RADCOR 2000}, hep-ph/0103040.
\bibitem{GH01} P.~Gambino and U.~Haisch, hep-ph/0109058.
\bibitem{PData00} Particle Data Group, The European Physical Journal {\bf C15} (2000) 1.
\bibitem{BLM01} C.~Bauer, M.~Luke and T.~Mannel, hep-ph/0102089.
\bibitem{borz2} F.~Borzumati, C.~Greub, T.~Hurth and D.~Wyler,
               Phys.\ Rev.\ {\bf D 62} (2000) 075005.
\bibitem{borz} F. Borzumati and C. Greub, \pr{58}{98}{074004}; 
               {\it ibid.} {\bf D59} (1999) 057501.
\bibitem{ciaf} P.~Ciafaloni, A.~Romanino and A.~Strumia,
                Nucl.\ Phys.\ {\bf B524} (1998) 361.
\bibitem{BMU00.2} C.~Bobeth, M.~Misiak and J.~Urban, \np{567}{2000}{153}.
\bibitem{Carena:2001uj} M.~Carena, D.~Garcia, U.~Nierste and C.~E.~Wagner,
               \pl{499}{2001}{141}. 
\bibitem{anlauf} H.~Anlauf, Nucl.\ Phys.\  {\bf B430} (1994) 245.
\bibitem{durham} P.~Gambino,
J.\ Phys.\ G {\bf G27} (2001) 1199.
\bibitem{Holzner} A.~Holzner, talk presented at the 
  {\it XXXVIth Rencontres de Moriond}, Les Arcs, March 2001.
\bibitem{taunu} The L3 Collaboration, \pl{396}{1997}{327}~ and references therein.
\bibitem{Xtaunu} The DELPHI Collaboration, \pl{496}{2000}{43}~ and references therein.
\bibitem{Moriond} F.~Blanc, talk presented at the 
  {\it XXXVIth Rencontres de Moriond}, Les Arcs, March 2001.
\bibitem{H00} A.H.~Hoang, hep-ph/0008102.
\bibitem{BSU97} I.~Bigi, M.~Shifman and N.~Uraltsev,
              Ann.\ Rev.\ Nucl.\ Part.\ Sci.\  {\bf 47} (1997) 591. 
\bibitem{BS99} M.~Beneke, \pl{434}{1998}{115};\\ 
               M.~Beneke and A.~Signer, \pl{471}{1999}{233}. 
\bibitem{ref_C_pert} N.~Cabibbo and L. Maiani, \pl{79}{1978}{109};\\
                     Y.~Nir, \pl{221}{1989}{184};\\
                     M.~Luke, M.J.~Savage and M.B.~Wise, \pl{345}{1995}{301};\\
                     A.~Czarnecki and K.~Melnikov, \pr{59}{1998}{014036};\\
                     T.~van~Ritbergen, \pl{454}{1999}{353}.
\bibitem{GBGS90} N.~Gray, D.J.~Broadhurst, W.~Grafe and K.~Schilcher,
                 \zpc{48}{1990}{673}.
\bibitem{LLMW99} Z.~Ligeti, M.~Luke, A.V.~Manohar and M.B.~Wise, \pr{60}{1999}{034019}.

\end{thebibliography}
\end{document}